\documentclass[%
reprint,
nofootinbib,
amsmath,
amssymb,
prd,
longbibliography,
superscriptaddress]{revtex4-2}

\usepackage{mathrsfs}
\usepackage{graphicx}
\usepackage{dcolumn}
\usepackage{bm}
\usepackage{lipsum}
\usepackage{graphics}
\usepackage{bm}
\usepackage{bbm}
\usepackage{color}
\usepackage{microtype}
\usepackage{IEEEtrantools}
\usepackage{mathrsfs}
\usepackage{amsthm}
\usepackage{enumitem}
\usepackage[normalem]{ulem}
\usepackage{hyperref}
\hypersetup{breaklinks=true}
\usepackage{graphics}
\usepackage[T1]{fontenc}
\usepackage{lmodern}

\definecolor{darkgreen}{rgb}{0,0.5,0}

\renewcommand{\P}{\mathcal{P}}

\newcommand{\res}{\mathcal{R}}

\newcommand{\scri}[0]{$\mathscr{I}^+$}
\newcommand{\order}[1]{\mathcal{O}(#1)}

\newcommand{\pkg}[1]{\textsc{#1}}

\newcommand{\beq}{\begin{equation}}
\newcommand{\eeq}{\end{equation}}

\font\ec=ecrm0800 at 12pt
\def\thorn{\hbox{\ec\char'336}}
\def\edth{\hbox{\ec\char'360}}

\def\mb{{\bar{m}}}

\renewcommand{\S}{{\mathcal S}}

\newcommand{\T}{{\mathcal T}}
\newcommand{\Lie}{{\mathcal L}}

\def\ddG{\delta^2G}
\def\dG{\delta G}
\def\ddR{\delta^2R}
\def\dR{\delta R}

\usepackage[T1]{fontenc}

\font\ec=ecrm0800 at 10pt
\def\thorn{\hbox{\ec\char'336}}
\def\edth{\hbox{\ec\char'360}}

\def\mb{{\bar{m}}}

\begin{document}

\title{Second-order Teukolsky formalism in Kerr spacetime: formulation and nonlinear source}

\author{Andrew Spiers}
\affiliation{%
 School of Mathematical Sciences \& School of Physics and Astronomy,
University of Nottingham, University Park, Nottingham, NG7 2RD, UK
}%
\affiliation{Nottingham Centre of Gravity, University of Nottingham, University Park, Nottingham, NG7 2RD, UK}
\affiliation{%
 School of Mathematical Sciences and STAG Research Centre, University of Southampton, Southampton, SO17 1BJ, United Kingdom
}
\author{Adam Pound}
\affiliation{%
 School of Mathematical Sciences and STAG Research Centre, University of Southampton, Southampton, SO17 1BJ, United Kingdom
}
\author{Jordan Moxon}
 \affiliation{TAPIR, California Institute of Technology, Pasadena, CA 91125, USA
}%

\date{\today}

\begin{abstract}
To fully exploit the capabilities of next-generation gravitational wave detectors, we need to significantly improve the accuracy of our models of gravitational-wave-emitting systems. This paper focuses on one way of doing so: by taking black hole perturbation theory to second perturbative order. Such calculations are critical for the development of nonlinear ringdown models and of gravitational self-force models of extreme-mass-ratio inspirals. In the most astrophysically realistic case of a Kerr background, a second-order Teukolsky equation presents the most viable avenue for calculating second-order perturbations. Motivated by this, we analyse two second-order Teukolsky formalisms and advocate for the one that is well-behaved for gravitational self-force calculations and which meshes naturally with recent metric reconstruction methods due to Green, Hollands, and Zimmerman [CQG 37, 075001 (2020)] and others. Our main result is an expression for the nonlinear source term in the second-order field equation; we make this available, along with other useful tools, in an accompanying \pkg{Mathematica} notebook. Using our expression for the source, we also show that infrared divergences at second order can be evaded by adopting a Bondi--Sachs gauge.

\end{abstract}

\maketitle


\section{\label{sec:level1} Introduction}

\subsection{Black hole perturbation theory beyond linear order}

Exact solutions of the Einstein field equations are few and far between. The most astrophysically relevant exact solutions, the Schwarzschild and Kerr spacetimes, describe the simplest systems of isolated, stationary bodies. To model dynamical spacetimes, particularly the types observable by gravitational wave detectors, one must generally resort to either numerically solving the fully nonlinear field equations~\cite{Baumgarte-Shapiro:NR} or using approximation methods~\cite{PoissonWill-book}. One such method, black hole perturbation theory~\cite{Pound-Wardell:2021}, approximates systems that closely resemble an isolated black hole. Since its inception, the development of this method has largely focused on perturbations at the leading, linear order. This has sufficed for most purposes; linear calculations are used in LIGO--Virgo--KAGRA templates~\cite{Taracchini-etal:2014,Bohe-etal:2016,Antonelli-etal:2019} to calibrate models of the final merger and subsequent ringdown at the end of an inspiral, for example. 

However, in the future, perturbative models must be substantially improved. Next-generation detectors will have improved sensitivity and broader frequency coverage, allowing them to observe a wider variety of systems with greater precision, but only if the accuracy of theoretical models keeps pace with detector technology~\cite{Purrer-Haster:2019}. Meeting the needs of next-generation detectors requires going to nonlinear perturbative orders~\cite{hindereretal2008,Pound-Wardell:2021,Loutrel-etal:2020}. In this paper, we take a step toward that goal, presenting new tools in second-order perturbation theory.

We specifically focus on the astrophysically realistic case of a perturbed Kerr black hole. This is expected to universally describe the final stage of a black hole binary, when the merged black holes ring down to a stationary Kerr state. It can also be remarkably accurate in describing the merger itself~\cite{McWilliams:2018,Rifat:2019ltp}. Recent work on second-order perturbations of Kerr has been motivated by these facts, particularly the desire for improved models of the final ringdown and the possible signatures of nonlinearity it may contain~\cite{green2020teukolsky,Loutrel-etal:2020,Ripley-etal:2020,Cheung:2022rbm,Sberna:2021eui,Mitman:2022qdl,Green:2022htq}.  
Here we are motivated more by another important source of gravitational waves: \textit{extreme-mass-ratio inspirals} (EMRIs). 
These systems occur in galactic nuclei when a stellar object of mass $m\sim1$--$ 100M_\odot$, such as a black hole or neutron star, slowly spirals into a massive black hole of mass $M\sim 10^4\text{--}10^7 M_\odot$~\cite{Amaro-Seoane:2020}. Due to the disparity between the two masses, the system is accurately described as a Kerr black hole perturbed by the orbiting companion~\cite{barackpound18}.

EMRIs are expected to be one of the premier sources for the space-based gravitational-wave detector LISA~\cite{Seoane-etal:2013,Babak-etal:2017,Berry-etal:2019}. 
While the EMRI event rate is uncertain, current estimates suggest that hundreds of detectable EMRIs will likely occur during the LISA mission's lifetime~\cite{Babak-etal:2017}. Each EMRI signal will be in the LISA band for $\sim 10^4$--$10^5$ wave cycles, with the companion spending most or all of that time in the strong-field region within 10 Schwarzschild radii of the central black hole. The emitted waveforms will therefore carry unique information about the strong-field regime of general relativity and the astrophysics of massive black holes. For a review of the stringent tests of general relativity, astrophysical measurements, and precise parameter extraction possible with EMRI observations, see~\cite{Amaro-Seoane-etal:2014,Babak-etal:2017,Berry-etal:2019,barausse2020LISAprospectsTests,LISA:2022yao,LISA:2022kgy}.

Most of these tests will only be possible using matched filtering with highly accurate waveform templates that maintain phase coherence with the signal for its full $\sim 10^5$ cycles. Currently, the only viable path to achieving such accuracy is using \textit{gravitational self-force} (GSF) theory~\cite{barackpound18,Pound-Wardell:2021}. In the context of a binary, this approach corresponds to an expansion in the binary's mass ratio $\varepsilon:=m/M\ll1$. At leading order in the expansion, the companion behaves as a test mass, following a geodesic of the central black hole geometry. At subleading orders, the perturbation produced by the companion influences the companion's own motion, effectively exerting a GSF on it. 

On the long timescale of an inspiral, the GSF's dominant effects are dissipative. It has been known for some time that to achieve the necessary phase accuracy, the GSF method must include {\em second-order} dissipative effects~\cite{rosenthal2006secondorder}. Ref.~\cite{hindereretal2008} showed this with a rigorous scaling argument, establishing that on the characteristic time $t\sim M/\varepsilon$ over which the orbit inspirals, the gravitational-wave phase has the following post-adiabatic expansion in  $\varepsilon$:
\begin{align}\label{eq:frequencies}
\varphi(t, \varepsilon)=\frac{1}{\varepsilon}\varphi^{(0)}(\varepsilon t) + \varphi^{(1)}(\varepsilon t) +\mathcal{O}(\varepsilon),
\end{align}
where the ``adiabatic'' (0PA) term $\varphi^{(0)}$ is dependent on the dissipative piece of the first-order GSF, and the ``first post-adiabatic'' (1PA) term $\varphi^{(1)}$ is dependent on the dissipative piece of the second-order GSF and the entire first-order GSF~\cite{hindereretal2008,Pound-Wardell:2021}. Hence, to ensure that cumulative phase errors remain small over the inspiral, we must know the dissipative piece of the second-order GSF. In fact, existing work on the conservative piece of the first-order GSF, which has been the focus of the GSF community, will likely not be of use in LISA data analysis if the dissipative second-order GSF is not also included, as they have comparable impacts on the phase evolution. 

Going to second order in perturbation theory brings a new set of challenges. While the general formalism for $n$th-order perturbations of generic background spacetimes is well understood~\cite{tomita1974non-linear,tomita1976nonlinear,bruni1997perturbations,sopuerta2004nonlinear}, concrete methods in particular spacetimes of interest are less well developed. Much of the work in nonlinear perturbation theory has focused on perturbations of cosmological spacetimes~\cite{mukhanov1997backreaction,matarrese1998relativistic,acquaviva2003gauge,tomita2005relativistic,nakamura2007second,McFadden-Skenderis:2011,pitrou2013xpand,uggla2014simple,bertacca2015galaxy} or of flat spacetime, either in post-Minkowskian  or post-Newtonian contexts~\cite{blanchet1986radiative,blanchet2014gravitational,PoissonWill-book}, where calculations have been carried to high orders~\cite{blanchet2014gravitational,Bern-etal:2021,Dlapa:2023hsl,Blanchet:2023bwj}. 
In the context of black hole perturbation theory, there are practical formulations of second-order perturbations of Schwarzschild spacetime, along with concrete calculations in some specific scenarios~\cite{poisson2010geometry,gleiser1996second,gleiser1996colliding,gleiser2000gravitational,nicasio2000second,brizuela2006second,ioka2007second,nakano2007second,brizuela2009complete,brizuela2010high,yang2015coupled,pound2020second,Miller-Pound:2020}. But much of this work has been restricted to vacuum perturbations and excluded certain perturbation modes. In the astrophysically realistic case of a spinning, Kerr black hole, substantially less has been done. After an early formulation by Campanelli and Lousto~\cite{l-c}, there has been very little work on the subject prior to a recent spate of papers by Green, Hollands, and Zimmerman (GHZ)~\cite{green2020teukolsky} and Loutrel and collaborators~\cite{Loutrel-etal:2020,Ripley-etal:2020}. 

Several key features of linear perturbation theory in Kerr do not extend to second order. At first order, the Teukolsky equation~\cite{teuk1972,teuk1973} enables one to solve a single, fully separable field equation for a gauge-invariant variable (either of the Weyl scalars $\psi_0$ or $\psi_4$). In vacuum regions, the Weyl scalar carries almost the entire information about the metric~\cite{Wald:1973}, and there is also a well-developed method of reconstructing the metric perturbation from the Weyl scalar due to Chrzanowski~\cite{chrzanowski1975vector} and Cohen and Kegeles~\cite{cohen1975space} (CCK). At second order, the Weyl scalars are sourced by nonlinear combinations of the full first-order metric perturbation; they are no longer invariant~\cite{l-c}; and even in vacuum regions, obtaining the second-order metric perturbation from them is highly nontrivial~\cite{green2020teukolsky, Toomani:2021jlo}.

GSF theory in this setting comes with more distinct difficulties. There is now a large body of work on second-order GSF theory in generic vacuum backgrounds~\cite{rosenthal2006secondorder,pound2010self,detweiler2012second-order,pound2012second,gralla2012,pound2012field,pound2014practical,pound2015,pound2015gaugeandmotion,pound2017HighlyRegularGauge,Upton}. 
A complete computational framework has been developed for the special case of quasicircular orbits in Schwarzschild spacetime~\cite{Wardell-Warburton:2015,Miller-Pound:2020,coupling,Miller-Pound:2021}, 
and concrete results have been obtained in that restricted case~\cite{pound2020second,Warburton:2021kwk,Wardell:2021fyy}. However, astrophysically realistic EMRI models will require such calculations in a Kerr background, as massive black holes are expected to have significant spin.

Currently, no second-order GSF calculations have been performed in a Kerr background, and methods for such a calculation are fledgling. In this paper, along with two sequel papers~\cite{paperII,paperIII}, we begin to develop a framework for these calculations. The core of our formulation, and the focus of the present paper, is a linear equation for a certain second-order Teukolsky variable. This equation has appeared previously~\cite{Spiers:GR22,green2020teukolsky}, and this paper partly serves simply to flesh out the basic analyses that were summarized in Ref.~\cite{Spiers:GR22}. But no reference, to our knowledge, has presented an explicit expression for the nonlinear source term in this second-order field equation, which is constructed from quadratic combinations of the first-order metric perturbation. Here we provide that expression, along with tools for working with it, in a supplemental \pkg{Mathematica} notebook~\cite{2nd-order-notebook}. The notebook was built upon an existing notebook, Ref.~\cite{npnotebook}, and uses the tensor-calculus package xAct~\cite{xact} (including the sub-packages xPert~\cite{brizuela2009xpert} and Spinors~\cite{gomez2012spinors}).  We also outline how this second-order Teukolsky formalism can be used in GSF calculations and some advantages it has over the earlier Campanelli--Lousto formulation that involved a slightly different field variable.

We expect our formalism to provide the basis for future second-order GSF calculations, and our discussion focuses on that application. However, our techniques are also applicable to any other second-order perturbative calculation in Kerr spacetime, particularly to nonlinear ringdown calculations.

\subsection{Outline}

We begin in Sec.~\ref{sec:Kerr-BHPT} with a review of perturbation theory and GSF theory in Kerr. We include a large portion of review material partly because many readers might be unfamiliar with second-order perturbation theory and partly as a necessary guide to the accompanying \pkg{Mathematica} notebook~\cite{2nd-order-notebook}. 

Sections~\ref{sec:second-order-scheme I} and \ref{sec:pure-2nd-Teuk} then examine two second-order Teukolsky formulations. We first review Campanelli and Lousto's second-order Teukolsky equation~\cite{l-c} and examine its utility for second-order GSF calculations. We show it generically has an ill-defined source term, but we suggest two potential ways around that. We then explore an alternative that we call the \textit{reduced} second-order Teukolsky equation. Unlike the Campanelli--Lousto equation, the reduced equation has a distributionally well-defined source term in the GSF problem. We also describe how the reduced formulation dovetails with the GHZ metric-reconstruction formalism~\cite{green2020teukolsky} (as well as other recently developed reconstruction methods~\cite{Aksteiner:2016pjt,Dolan:2021ijg,Toomani:Lorenz,Wardell:Capra2022,Dolan:Capra2022}) and how both connect with the Teukolsky puncture/effective-source scheme developed in Ref.~\cite{Toomani:2021jlo}.

In Sec.~\ref{sec:asymptotics}, we next examine the asymptotic properties of the source of the reduced second-order Teukolsky equation and comment on how the infrared divergences that generically arise in second-order GSF calculations~\cite{PoundLargeScales} can be avoided by imposing Bondi-type gauge conditions. One of the sequel papers~\cite{paperII} will describe how to incorporate those gauge conditions into our framework.

We conclude in Sec.~\ref{sec:conclusion} by summarizing how this paper sets the stage for several followups, including the two sequels mentioned above as well as a framework and implementation specialized to Schwarzschild spacetime~\cite{coupling,2ndOrderTeukolskyInSchwarzschild}.

Throughout this paper, we use lowercase Latin letters to denote abstract indices and Greek letters $\alpha,\beta\in\{0,1,2,3\}$ to denote components in a coordinate basis. Greek indices in square brackets denote components in a tetrad basis. We adopt the $(-+++)$ metric signature and geometric units with $G=c=1$. $(t,r,\theta,\phi)$ denote Boyer-Lindquist coordinates.

\section{\label{sec:Kerr-BHPT}Black hole perturbation theory}

In this summary of black hole perturbation theory, we review the generic structure of perturbation theory in general relativity; Newman--Penrose (NP), first-order Teukolsky, and metric-reconstruction formalisms; and the applications of these things in GSF theory. We refer to Refs.~\cite{Pound-Wardell:2021,barackpound18} for a more thorough review. Readers familiar with this review material can skip freely to Sec.~\ref{sec:second-order-scheme I}.

\subsection{Perturbation theory in general relativity}\label{sec:generic perturbation theory}

We consider a one-parameter family of spacetimes with metrics $g_{ab}(\varepsilon)$. In the bulk of the paper $\varepsilon$ may be any small parameter, although we will take it to be the mass ratio when specializing to a binary system.

We assume tensors  $A_{a_1...a_m}^{b_1...b_n}$ on this family admit series expansions in powers of $\varepsilon$,
\begin{align}\label{eq:generic expansion}
	A_{a_1\cdots a_m}^{b_1\cdots b_n}&=A_{a_1\cdots a_m}^{(0)b_1\cdots b_n}+\varepsilon A_{a_1\cdots a_m}^{(1)b_1\cdots b_n} \notag \\ & \ \ \ +\varepsilon^2 A_{a_1\cdots a_m}^{(2)b_1\cdots b_n} +\ldots
\end{align}
In particular, we expand the metric and stress-energy tensor as
\begin{align}\label{eq:metricexpansion}
g_{ab} &= g^{(0)}_{ab}+\varepsilon h^{(1)}_{ab} +\varepsilon^2 h^{(2)}_{ab} +\ldots+ \varepsilon^n h^{(n)}_{ab}+\ldots,\\
T_{ab} &= T^{(0)}_{ab}+\varepsilon T^{(1)}_{ab} +\varepsilon^2 T^{(2)}_{ab} +\ldots+ \varepsilon^n T^{(n)}_{ab}+\ldots
\end{align}
We then expand the Einstein equation,
\begin{align}\label{eq:EFE}
G_{ab}[g_{ab}]=8\pi T_{ab},
\end{align}
to obtain field equations for the successive terms in Eq.~\eqref{eq:metricexpansion}. 

The structure of the field equations is made clearer by first expanding the Einstein tensor in powers of the total metric perturbation $h_{ab}:=g_{ab} - g^{(0)}_{ab}$:
\begin{multline}\label{eq:G expansion}
G_{ab}[g_{ab}] = G_{ab}[g^{(0)}_{ab}] + \delta G_{ab}[h_{ab}] +\delta^2G_{ab}[h_{ab},h_{ab}] \\
+\ldots +  \delta^n G_{ab}[h_{ab},\ldots,h_{ab}] +\ldots,
\end{multline}
where $\delta G_{ab}$ is the linearised Einstein tensor, $\delta^2G_{ab}$ is quadratic in $h_{ab}$, and so on; each of the operators $\delta^n G_{ab}$ is linear in each of its $n$ arguments. We omit the textbook expression for $\delta G_{ab}$, but for easy reference, we include here the expression for the quadratic Ricci tensor,
\begin{align}\label{eq:d2Rab}
\delta^2 R_{ab} &=-\frac{1}{2}\bar h^{cd}{}_{;d} (2h_{c(a;b)}-h_{ab;c}) \notag \\ &\quad +\frac{1}{4}h^{cd}{}_{;a} h_{cd;b}  + \frac{1}{2}h^{c}{}_{b}{}^{;d}(h_{ca;d}-h_{da;c}) \notag \\ &\quad  -\frac{1}{2}h^{cd}(2h_{c(a;b)d}-h_{ab;cd}-h_{cd;ab}).
\end{align}
$\delta^2G_{ab}$ then comprises the quadratic terms in the trace reversal of $R_{ab}$,
\begin{multline}\label{eq:d2Gab}
\delta^2 G_{ab} = \delta^2 R_{ab} -\frac{1}{2}g^{(0)}_{ab}g^{(0)cd}\delta^2 R_{cd} \\
- \frac{1}{2}\left(h_{ab}g^{(0)cd}-g^{(0)}_{ab}h^{cd}\right)\dR_{cd}.
\end{multline}
Here indices are raised and lowered with the background metric, a semicolon denotes covariant differentiation compatible with the background metric, and $\bar h_{ab}:=h_{ab}-\frac{1}{2}g^{(0)}_{ab}g^{(0)cd}h_{cd}$.

If we substitute the expansion of $h_{ab}$ in powers of $\varepsilon$ into Eq.~\eqref{eq:G expansion}, the Einstein equation becomes
\begin{align}\label{eq:expandedEFE}
G_{ab}[g^{(0)}_{ab}] &+\varepsilon\delta G_{ab}[h^{(1)}_{ab}] + \varepsilon^2\left(\delta G_{ab}[h^{(2)}_{ab}]+\delta^2G_{ab}[h^{(1)}_{ab}] \right) \notag \\
&= 8\pi \left(T^{(0)}_{ab}+\varepsilon T^{(1)}_{ab}+\varepsilon^2 T^{(2)}_{ab}\right)+\mathcal{O}(\varepsilon^3). 
\end{align}
Equating coefficients of powers of $\varepsilon$ then yields a nonlinear equation for the background metric, $G_{ab}[g^{(0)}_{ab}]=8\pi T^{(0)}_{ab}$, together with a sequence of \emph{linear} equations for the perturbations $h^{(n)}_{ab}$,
\begin{align} 
\delta G_{ab}[h^{(1)}_{ab}]&=8\pi T^{(1)}_{ab},\label{eq:linearEFE}\\
\delta G_{ab}[h^{(2)}_{ab}]&=8\pi T^{(2)}_{ab}-\delta^2G_{ab}[h^{(1)}_{ab},h^{(1)}_{ab}],\label{eq:quadraticEFE}\\
\delta G_{ab}[h^{(3)}_{ab}]&=\ldots
\end{align}
For a vacuum background such as Kerr, we have 
\beq
G_{ab}[g^{(0)}_{ab}]=T^{(0)}_{ab}=0. 
\eeq
In principle, perturbation theory then boils down to solving the sequence of linear equations for each successive $h^{(n)}_{ab}$ (together with any equations governing the matter fields in the system). These equations have the same left-hand side at every order; their only difference is their source terms on the right-hand side, which involve nonlinear combinations of lower-order metric perturbations. 

While conceptually simple, solving these linear equations in a Kerr background is challenging (and numerically expensive) because each of them comprises a non-separable set of coupled partial differential equations.\footnote{As a consequence of Kerr's axisymmetry and stationarity, one can \emph{partially} separate the equations by expanding the components $h^{(n)}_{\alpha\beta}$ in modes $h^{(nm\omega)}_{\alpha\beta}(r,\theta)e^{i(m\phi-\omega t)}$, leading to two-dimensional elliptic equations for each of the coefficients $h^{(nm\omega)}_{\alpha\beta}$. This route is being actively explored~\cite{Kerr2D}.} Given this challenge, instead of tackling the equations directly, at first order, one generally solves the linearized Einstein equation indirectly by reconstructing $h^{(1)}_{ab}$ from a single complex scalar that satisfies the (fully separable) Teukolsky equation. We review these metric reconstruction methods in Sec.~\ref{sec:reconstruction}. The derivations of the Teukolsky equation and metric reconstruction generally rely on using the NP formalism, which we review next.

\subsection{Newman--Penrose formalism}\label{subsec:NP}

The NP formalism utilizes an orthonormal basis of null vectors, 
\begin{align}
e_{[\mu]}^a=\{e_{[1]}^a,e_{[2]}^a,e_{[3]}^a,e_{[4]}^a\} := \{l^a,n^a,m^a,\bar{m}^a\},
\end{align}
chosen such that $l^a$ and $n^a$ are real and $m^a$ is complex (with a bar denoting complex conjugation). They satisfy the orthonormality conditions
\begin{equation}
l^an_a=-1,\quad m^a\bar{m}_a=1,
\end{equation}
and $g_{ab}e^a_{[\mu]}e^b_{[\nu]}=0$ for all other combinations of tetrad legs. The metric in this basis therefore reads
\begin{equation}\label{eq:NPg}
  g_{ab} = -2 l_{(a} n_{b)} +2 m_{(a} \bar{m}_{b)} ,
\end{equation}
where parentheses denote symmetrization. 

In the NP formalism, the connection is represented using Ricci rotation coefficients,
\begin{equation}
\gamma_{[\mu][\alpha][\beta]}:=e_{[\mu]}^ke_{[\alpha]k;i}e_{[\beta]}^i.
\end{equation}
The independent components of the Ricci rotation coefficients are denoted using complex scalars,
\begin{align}
\kappa=-\gamma_{[3][1][1]},\ \tau=-&\gamma_{[3][1][2]},  \ \sigma=-\gamma_{[3][1][3]}, \notag
\\  \
\rho=-\gamma_{[3][1][4]},  
 \ \pi=-&\gamma_{[2][4][1]},\ \nu=-\gamma_{[2][4][2]}, \notag
\\   \ \mu=-\gamma_{[2][4][3]}&, \ \lambda=-\gamma_{[2][4][4]}, \notag
\\
\epsilon=-\frac{\gamma_{[2][1][1]}+\gamma_{[3][4][1]}}{2}&,  \ \gamma=-\frac{\gamma_{[2][1][2]}+\gamma_{[3][4][2]}}{2}, \notag
\\ 
\beta=-\frac{\gamma_{[2][1][3]}+\gamma_{[3][4][3]}}{2}&,  \ \alpha=-\frac{\gamma_{[2][1][4]}+\gamma_{[3][4][4]}}{2},
\end{align}
known as \textit{spin coefficients}. 

The covariant derivatives along tetrad vectors are written as operators $\bm{D},\bm{\Delta}, \bm{\delta}$, and $\bar{\bm{\delta}}$, which are defined as
\begin{align}\label{paradervs}
\bm{D}:=l^a \nabla_a, \ \bm{\Delta}:=n^a \nabla_a, \ &\bm{\delta}:=m^a \nabla_a, \ \bar{\bm{\delta}}:=\bar{m}^a \nabla_a.
\end{align}
We use boldface symbols for these directional derivatives to distinguish them from $\delta$ (denoting functional derivatives as in the preceding section) and $\Delta$ (denoting a gauge transformation as in the next section). It will be useful to also define
\begin{align}
\bar{d}_3 &:=\bar{\bm{\delta}}-\bar{\tau}+\bar{\beta} +3\alpha+4\pi,\label{eq:d3}\\
\bar{d}_4 &:=\bm{\Delta}+3\gamma-\bar{\gamma}+4\mu+\bar{\mu},\label{eq:d4}
\end{align}
following the notation of Ref.~\cite{l-c}.

In the NP formalism, one expresses the vacuum curvature by defining Weyl scalars, five complex scalars that represent the ten degrees of freedom of the Weyl tensor $C_{abcd}$:
\begin{align}
\psi_0 &= C_{abcd}l^a m^b l^c m^d, \label{eq:psi0 def}\\
\psi_1 &= C_{abcd}l^a m^b l^c n^d, \\
\psi_2 &= C_{abcd}l^a m^b \bar m^c n^d,\\
\psi_3 &= C_{abcd}l^a n^b \bar m^c n^d, \\
\psi_4 &= C_{abcd}n^a\bar m^b n^c \bar m^d. \label{eq:psi4 def}
\end{align}

All of the above is generic; it applies for any metric, whether the background or the exact spacetime. For our one-parameter family of spacetimes, each NP quantity is expanded in powers of $\varepsilon$, starting with the tetrad itself:
\begin{equation}\label{eq:tetrad expansion}
  e^a_{[\mu]} \to e^{(0)a}_{[\mu]} + \varepsilon e^{(1)a}_{[\mu]} + \mathcal{O}(\varepsilon^2).
\end{equation}
When using the NP formalism to describe perturbed quantities, we will only rarely write any exact quantity. We, therefore, simplify the notation by dropping the $(0)$ label on background quantities. For example, rather than using $e^a_{[\mu]}$ to denote the exact tetrad, we use it to denote $e^{(0)a}_{[\mu]}$; this is the reason for the arrow, instead of an equality, in Eq.~\eqref{eq:tetrad expansion}. We express the first-order tetrad perturbations and spin coefficients in terms of $h^{(1)}_{ab}$ in Appendix~\ref{app:1st-order-NP} (while noting that such expressions are necessarily not unique because of the freedom to rotate the tetrad, reviewed below).

If the zeroth-order tetrad legs $l^a$ and $n^a$ are chosen to lie along the principal null directions of Kerr, four of the Weyl scalars and four of the NP spin coefficients are made to vanish at zeroth order~\cite{chandrabook}:
\begin{align}\label{petrovconds}
&\psi_0=0,\ \psi_1=0,\ \psi_3=0,\ \psi_4=0 \\
&\kappa=0,\ \lambda=0,\ \nu=0,\ \sigma=0.
\end{align}
We will always make this choice, aligning $l^a$ with the outgoing principle null direction and $n^a$ with the ingoing one. $\epsilon$ can also be made to vanish by further specializing to the Kinnersley tetrad~\cite{kinnersley1969type}. 

Beyond zeroth order, our main quantities of interest are the perturbations of $\psi_0$ and $\psi_4$. At linear order, these are independent of the tetrad perturbations; this follows from the fact that $C_{abcd}\delta(l^a m^b l^c m^d) = 0 = C_{abcd}\delta(n^a \bar m^b n^c \bar m^d)$~\cite{Bishop:2016lgv}, such that the definitions~\eqref{eq:psi0 def} and \eqref{eq:psi4 def} imply
\beq\label{eq:deltapsi}
\psi^{(1)}_{4} = \delta\psi_4[h^{(1)}_{ab}] = \delta C_{abcd}[h^{(1)}_{ab}]n^a \mb^b n^c \mb^d
\eeq
and analogously for $\psi^{(1)}_{0}$. Appealing to the same identities, we see that the second-order analogue of Eq.~\eqref{eq:deltapsi} has the more complicated structure
\begin{equation}\label{eq:psi42}
\psi^{(2)}_4 = \delta \psi_4[h^{(2)}_{ab}] + \delta^2\psi_4\bigl[h^{(1)}_{ab},e^{(1)a}_{[\mu]}\bigr],
\end{equation}
where the second term is quadratic in $h^{(1)}_{ab}$ and $e^{(1)a}_{[\mu]}$ (and is not bilinear in its two arguments). This structure of $\psi^{(2)}_4$ will be important for our discussions in Sec.~\eqref{sec:pure-2nd-Teuk}.

For later reference, we write the linear operator in Eqs.~\eqref{eq:deltapsi} and \eqref{eq:psi42} as
\beq
\mathcal{T}:=\delta\psi_4[\cdot]
\eeq
and include here its explicit NP form,
\begin{align}
    \T^{ab} &= -\frac12 \biggl\{ (\bar{\bm{\delta}} -\bar\tau +3\alpha + \bar\beta)(\bar{\bm{\delta}} -\bar\tau +2\alpha +2\bar\beta)n^an^b \notag \\
    &\quad +(\bm{\Delta} +\bar\mu +3\gamma-\bar\gamma)(\bm{\Delta} +\bar\mu +2\gamma-2\bar\gamma)\mb^a\mb^b     \notag \\
    &\quad -\Bigl[ (\bm{\Delta} +\bar\mu +3\gamma-\bar\gamma)(\bar{\bm{\delta}} -2\bar\tau +2\alpha) \notag \\
    &\quad +(\bar{\bm{\delta}} -\bar\tau +3\alpha + \bar\beta)(\bm{\Delta} +2\bar\mu +2\gamma) \Bigr] n^{(a}\mb^{b)}
    \biggr\}.\!\!\label{eq:T}
\end{align}
Following Ref.~\cite{l-c}, we also write Eq.~\eqref{eq:psi42} as
\begin{align}\label{eq:psi42deconstructed}
\psi_4^{(2)}&=\psi_{4L}^{(2)}+\psi_{4Q}^{(2)},
\end{align}
where $\psi_{4L}^{(2)}:=\mathcal{T}[h^{(1)}_{ab}]$ and $\psi_{4Q}^{(2)}:=\delta^2\psi_4[h^{(1)}_{ab},e^{(1)a}_{[\mu]}]$. $\psi_{4Q}^{(2)}$ is given in NP form in Eq.~(B3) of Ref.~\cite{l-c} as well as in the supplemental \pkg{Mathematica} notebook~\cite{2nd-order-notebook}. 


In this section and throughout the body of the paper, we use standard NP quantities. The accompanying \pkg{Mathematica} notebook~\cite{2nd-order-notebook} also presents results in terms of the refined NP formalism due to Geroch, Held, and Penrose (GHP)~\cite{geroch1973space}. We summarize that reformulation in Appendix~\ref{app:GHP}.

\subsection{Gauge transformations and infinitesimal tetrad rotations}\label{subsec:gauge}

Perturbation theory in the NP formalism has two types of perturbative gauge freedom: one corresponding to changes in the identification between points in the exact spacetime and points in the background; and one corresponding to near-identity rotations of the tetrad. To avoid ambiguity, we refer to a change of identification as a gauge transformation, and we refer to a change of tetrad as an infinitesimal tetrad rotation.

Gauge transformations are equivalent to near-identity coordinate transformations. For a generic expansion of the form~\eqref{eq:generic expansion}, a gauge transformation leaves the zeroth-order term unchanged while altering the subleading terms as $A_{a_1\cdots a_m}^{(n)b_1\cdots b_n}\to A_{a_1\cdots a_m}^{(n)b_1\cdots b_n}+\Delta A_{a_1\cdots a_m}^{(n)b_1\cdots b_n}$, where%
\begin{align}
\Delta A_{a_1\cdots a_m}^{(1)b_1\cdots b_n} &= \Lie_{\xi^c_{(1)}} A_{a_1\cdots a_m}^{(0)b_1\cdots b_n},\label{DeltaA1}\\
\Delta A_{a_1\cdots a_m}^{(2)b_1\cdots b_n} &= \Lie_{\xi^c_{(2)}} A_{a_1\cdots a_m}^{(0)b_1\cdots b_n} + \frac{1}{2}\Lie_{\xi^c_{(1)}}\Lie_{\xi^c_{(1)}}A_{a_1\cdots a_m}^{(0)b_1\cdots b_n}\nonumber\\
&\quad + \Lie_{\xi^c_{(1)}}A_{a_1\cdots a_m}^{(1)b_1\cdots b_n}.\label{DeltaA2}
\end{align}
For a derivation of this, see Ref.~\cite{bruni1997perturbations} or Sec. IVA of Ref.~\cite{pound2015gaugeandmotion}. Here $\Lie$ denotes a Lie derivative, and the vector fields $\xi^a_{(n)}$ are referred to as the generators of the transformation.

Applied to the perturbations $h^{(n)}_{ab}$ and $T^{(n)}_{ab}$, the general rules~\eqref{DeltaA1} and \eqref{DeltaA2} become%
\begin{align}
\Delta h^{(1)}_{ab} &= \Lie_{\xi^c_{(1)}} g^{(0)}_{ab},\label{Deltah1}\\
\Delta h^{(2)}_{ab} &= \Lie_{\xi^c_{(2)}} g^{(0)}_{ab} + \frac{1}{2}\Lie^2_{\xi^c_{(1)}}g^{(0)}_{ab} + \Lie_{\xi^c_{(1)}} h^{(1)}_{ab},\label{Deltah2}
\end{align}
and
\begin{align}
\Delta T^{(1)}_{ab}&=0,\label{Delta T1}\\
\Delta T^{(2)}_{ab} &= \Lie_{\xi^c_{(1)}}T^{(1)}_{ab},\label{Delta T2}
\end{align}
where we have specialized to a vacuum background with $T^{(0)}_{ab}=0$.

The perturbative Einstein equations~\eqref{eq:linearEFE} and \eqref{eq:quadraticEFE} (and their analogues at all higher orders) are invariant under a generic gauge transformation. This is easily seen by moving all curvature terms to the left-hand side while noting $\dG_{ab}[h^{(1)}_{ab}]=G^{(1)}_{ab}$ and $(\dG_{ab}[h^{(1)}_{ab}]+\ddG_{ab}[h^{(1)}_{ab},h^{(1)}_{ab}])=G^{(2)}_{ab}$. The invariance then follows from Eqs.~\eqref{DeltaA1} and \eqref{DeltaA2}, so long as the field equations are satisfied in the original gauge. For example, at first order:
\begin{equation}
\Delta G^{(1)}_{ab} = {\cal L}_{\xi^c_{(1)}}G^{(0)}_{ab} = 8\pi{\cal L}_{\xi^c_{(1)}}T^{(0)}_{ab} = 8\pi\Delta T^{(1)}_{ab}.
\end{equation}
In a vacuum background, we also have the stronger statement that each side of the first-order Einstein equation~\eqref{eq:linearEFE} is separately invariant; this follows immediately from the vanishing of the zeroth-order fields $G^{(0)}_{ab}$ and $T^{(0)}_{ab}$. Because $\Delta G^{(1)}_{ab}$ is also given by $\dG_{ab}[\Delta h^{(1)}_{ab}]=\dG_{ab}[{\cal L}_{\xi^c}g^{(0)}_{ab}]$, its vanishing also implies the standard identity
\beq
\dG_{ab}[{\cal L}_{\xi^c}g^{(0)}_{ab}]= 0\label{Delta dG1}
\eeq
for any $\xi^c$. But at second order, while the Einstein equation~\eqref{eq:quadraticEFE} is still invariant, each separate side is \emph{not}. Using Eqs.~\eqref{Deltah2} and \eqref{Delta dG1}, we see the left-hand side transforms as
\begin{align}
\Delta \dG_{ab}[h^{(2)}_{ab}] &= \dG_{ab}[\Delta h^{(2)}_{ab}] \nonumber\\
&= \dG_{ab}\!\left[\tfrac{1}{2}\Lie_{\xi^c_{(1)}}\Lie_{\xi^c_{(1)}}g^{(0)}_{ab} + \Lie_{\xi^c_{(1)}} h^{(1)}_{ab}\right].\label{Delta dG2}
\end{align}
From the invariance of the equation as a whole,
the right-hand side transforms in the same way,
\begin{equation}
\Delta\left(8\pi T^{(2)}_{ab}-\ddG_{ab}[h^{(1)}_{ab}]\right) = \dG_{ab}[\Delta h^{(2)}_{ab}].
\end{equation}
This gauge dependence will be important in later sections.

We now turn to infinitesimal tetrad rotations, which correspond to near-identity Lorentz transformations (boosts or spatial rotations) of the tetrad legs. Under such a transformation, the zeroth-order legs are unchanged, while the perturbations $e^{(n)a}_{[\mu]}$ are transformed. We will only require the first-order tetrad perturbations, which transform as 
\beq\label{eq:tetrad rotation}
e^{(1)a}_{[\mu]} \to e^{(1)a}_{[\mu]} + B_{[\mu]}{}^{[\nu]} e^{a}_{[\nu]},
\eeq
where $B_{[\mu][\nu]}$ is an arbitrary antisymmetric matrix. Here frame indices are raised with the inverse of the Minkowski metric $\eta_{[\mu][\nu]}=g^{(0)}_{ab}e^a_{[\mu]}e^b_{[\nu]}$. Following our convention discussed in the previous section, $e^{a}_{[\mu]}$ denotes the zeroth-order tetrad legs. 

In a Kerr background (except in the Schwarzschild limit), $\psi^{(n)}_{2}$ can be set to zero at all perturbative orders through a gauge transformation. At first order, for example, it transforms as
\beq
\Delta\psi^{(1)}_{2} = {\cal L}_{\xi^c_{(1)}}\psi_{2} = \xi^{r}_{(1)}\partial_r\psi_{2} + \xi^{\theta}_{(1)}\partial_\theta\psi_{2}; 
\eeq
since $\psi_{2}$ and $\psi^{(1)}_{2}$ are complex (except when the Kerr spin parameter vanishes), this implies we can set  $\psi^{(1)}_{2}$ to zero by solving $\Delta\psi^{(1)}_{2}=-\psi^{(1)}_{2}$ for the two components $\xi^{r}_{(1)}$ and $\xi^{\theta}_{(1)}$. We can likewise set $\psi^{(n)}_{2}=0$ through appropriate choices of  $\xi^{r}_{(n)}$ and $\xi^{\theta}_{(n)}$. Similarly, the Weyl scalars $\psi^{(n)}_{1}$ and $\psi^{(n)}_{3}$ can always be set to zero through an infinitesimal tetrad rotation. Therefore, at all orders, $\psi^{(n)}_{0}$ and $\psi^{(n)}_{4}$ are the only Weyl scalars that carry gauge- and tetrad-invariant information.

Moreover, at first order, the Weyl scalars $\psi^{(1)}_0$ and $\psi^{(1)}_4$ are both gauge and tetrad invariant.  They are trivially invariant under infinitesimal tetrad rotations because, from Eq.~\eqref{eq:deltapsi}, they are independent of the tetrad perturbations $e^{(1)a}_{[\mu]}$. They are trivially gauge invariant by virtue of Eq.~\eqref{DeltaA1} because the background scalars $\psi_0$ and $\psi_4$ vanish. Just as the invariance of $G^{(1)}_{ab}$ implied Eq.~\eqref{Delta dG1}, the gauge invariance of $\psi^{(1)}_0$ and $\psi^{(1)}_4$ implies
\beq\label{eq:Delta dpsi}
\delta\psi_0[{\cal L}_{\xi^c}g^{(0)}_{ab}] = 0 = \delta\psi_4[{\cal L}_{\xi^c}g^{(0)}_{ab}]
\eeq
for all vectors $\xi^c$.

On the other hand, the second-order perturbations $\psi^{(2)}_{0}$ and $\psi^{(2)}_4$ are neither gauge invariant nor infinitesimal-tetrad-rotation invariant~\cite{l-c}. We review their transformation properties in Sec.~\ref{subsec:LCT-transform} below.

\subsection{First-order Teukolsky equation}\label{sec:teukolskyeq}

In first-order perturbation theory, the invariant Weyl scalars $\psi^{(1)}_0$ and $\psi^{(1)}_4$ satisfy decoupled, fully separable Teukolsky equations~\cite{teuk1973, chandrabook}. These equations are most easily derived from the Penrose wave equation~\cite{Ryan:1974},\footnote{The original derivation due to Teukolsky instead began from a selection of perturbed Bianchi and Ricci identities in NP form~\cite{teuk1973,chandrabook}. That derivation is reproduced in the accompanying \pkg{Mathematica} notebook~\cite{2nd-order-notebook}.}
\beq\label{eq:Penrose wave equation}
\Box C_{abcd} - 4C_{aef[c}C_{d]}{}^f{}_{b}{}^e + C_{abef}C^{ef}{}_{cd} = S_{abcd},
\eeq
which is itself a consequence of the Ricci identity and Bianchi identities. Here $\Box:=g^{ab}\nabla_a\nabla_b$, $S_{abcd}$ is made up of terms involving the stress-energy tensor, and all quantities are constructed from the exact metric (as opposed to the background metric). Contracting Eq.~\eqref{eq:Penrose wave equation} with appropriate tetrad legs and linearizing, one finds the Teukolsky equations for $\psi^{(1)}_0$ and $\psi^{(1)}_4$. We write these equations concisely as 
\begin{align}
\mathcal{O}'\psi_0^{(1)} &= 8\pi\mathcal{S}'[T^{(1)}_{ab}],\label{eq:teuk-concise-prime}\\
\mathcal{O}\psi_4^{(1)} &= 8\pi\mathcal{S}[T^{(1)}_{ab}],\label{eq:teuk-concise}
\end{align}
where $\mathcal{O}$, $\mathcal{S}$, and their primed versions are second-order linear differential operators; the prime here denotes the GHP operation $(l^a \leftrightarrow n^a,m^a \leftrightarrow \bar{m}^a)$ explained in Appendix~\ref{app:GHP}. 

One can work with either of the variables $\psi_0^{(1)}$ or $\psi_4^{(1)}$. Here we focus on $\psi_4^{(1)}$, in which case the differential operators in the Teukolsky equation are given by
\begin{multline}\label{eq:O}
{\cal O} = (\bm{\Delta}  +3\gamma-\bar{\gamma}+4\mu+\bar{\mu})(\bm{D}+4\epsilon-\rho) \\ 
-(\bar{\bm{\delta}}-\bar{\tau}+\bar{\beta}+3\alpha+4\pi)(\bm{\delta}-\tau+4\beta)-3\psi_2
\end{multline}
and
\begin{align}
\mathcal{S}^{ab} &= \frac{1}{2}\bar{d}_4^{(0)}\Bigl[(\bar{\bm{\delta}}-2\bar{\tau}+2\alpha)n^{(a}\bar m^{b)} \notag\\
&\qquad\qquad - (\bm{\Delta}+2\gamma-2\bar{\gamma}+\bar{\mu})\bar{m}^{a} \bar{m}^{b}\Bigr]  \notag\\
&\quad +  \frac{1}{2}\bar{d}_3^{(0)} \Bigl[(\bm{\Delta}+2\gamma+2\bar{\mu})n^{(a} \bar m^{b)} \notag\\
&\qquad\qquad\quad -  (\bar{\bm{\delta}}-\bar{\tau}+2\bar{\beta}+2\alpha)n^a n^b \Bigr],\label{eq:S}
\end{align}
where all quantities here are zeroth order, and $\bar{d}_3^{(0)}$ and $\bar{d}_4^{(0)}$ are the zeroth-order  versions of the operators in Eqs.~\eqref{eq:d3} and \eqref{eq:d4}. We have restored indices on ${\cal S}$ to indicate that it acts on rank-2 tensors to return scalars (i.e., ${\cal S}[T_{ab}]:={\cal S}^{ab}T_{ab}$). 

Equations~\eqref{eq:teuk-concise-prime} and~\eqref{eq:teuk-concise} can be written in a common, separable form known as the Teukolsky master equation,
\beq\label{eq:masterTeuk s}
{}_s\hat{\cal O}{}_s\psi = {}_s S, 
\eeq
where the left subscript indicates spin weight, and where ${}_s\hat{\cal O}$ is given explicitly in Eq.~(4.7) of Ref.~\cite{teuk1973}. The relationship between the master scalars ${}_s\psi$ and the Weyl scalars depends on the choice of background tetrad~\cite{Pound-Wardell:2021}. In the Kinnersley tetrad, ${}_2\psi = \psi_0^{(1)}$, and Eq.~\eqref{eq:masterTeuk s} for $s=2$ is identical to Eq.~\eqref{eq:teuk-concise-prime}; while ${}_{-2}\psi = \rho^{-4}\psi_4^{(1)}$, and  Eq.~\eqref{eq:masterTeuk s} for $s=-2$ reads
\begin{align}\label{eq:masterTeuk}
{}_{-2}\hat{\mathcal{O}}[\rho^{-4}\psi_4^{(1)}]=16\pi\Sigma \rho^{-4}\mathcal{S}[T^{(1)}_{ab}],
\end{align}
where $\Sigma=r^2+a^2\cos^2\theta$ and ${}_{-2}\hat{\mathcal{O}}$ is related to $\mathcal{O}$ by ${}_{-2}\hat{\mathcal{O}} = 2\Sigma\rho^{-4}{\cal O}\rho^4$. The equations~\eqref{eq:teuk-concise-prime} and~\eqref{eq:teuk-concise} may not be manifestly separable, but Eq.~\eqref{eq:masterTeuk s} is immediately separable in a basis of spin-weighted spheroidal harmonics ${}_sS_{\ell m\omega}$. For example, Eq.~\eqref{eq:masterTeuk} is separated with the ansatz
\beq
\rho^{-4}\psi^{(1)}_4=\int d\omega\sum_{\ell m}{}_{-2}\psi^{(1)}_{\ell m \omega}(r){}_{-2}S_{\ell m\omega}(\theta)e^{im\phi-i\omega t}.\!\!
\eeq
This reduces Eq.~\eqref{eq:masterTeuk} to an ordinary differential equation for each radial coefficient  ${}_{-2}\psi^{(1)}_{\ell m\omega}$. 


\subsection{Metric reconstruction}\label{sec:reconstruction}

$\psi_0^{(1)}$ and $\psi_4^{(1)}$ directly encode the gravitational waves emitted to future null infinity~(\scri)~\cite{teuk1973} and into the primary black hole horizon~\cite{Teukolsky:1974yv}. Moreover, in vacuum, each of them contains \emph{all} the information about the metric perturbation $h^{(1)}_{ab}$ (up to trivial perturbations towards other Kerr solutions)~\cite{Wald:1973}. As recently shown by GHZ~\cite{green2020teukolsky}, even in nonvacuum $\psi_0^{(1)}$ and $\psi_4^{(1)}$ each contain ``most'' of the information in $h^{(1)}_{ab}$. Metric reconstruction methods are a realisation of this fact, allowing one to reconstruct any perturbation $h^{(1)}_{ab}$ from its corresponding $\psi_0^{(1)}$ or $\psi_4^{(1)}$ (along with a few other simple ingredients described below, and again excluding trivial perturbations). We again focus on schemes that begin from $\psi^{(1)}_4$.


For vacuum perturbations, the standard CCK reconstruction method~\cite{chrzanowski1975vector, cohen1975space, kegeles1979constructive}  
begins with Wald's operator identity~\cite{wald1978}
\begin{align}\label{eq:Wald-OI}
\mathcal{O}\mathcal{T}=\mathcal{S}\mathcal{E},
\end{align}
where $\mathcal{O}$, $\mathcal{T}$, and ${\cal S}$ are the operators defined in Eqs.~\eqref{eq:O}, \eqref{eq:T}, and~\eqref{eq:S}, and $\mathcal{E}$ is the linearized Einstein operator,
\beq
\mathcal{E}:=\delta G_{ab}[\cdot].
\eeq
Equation~\eqref{eq:Wald-OI} is simply the statement that the linear Teukolsky equation~\eqref{eq:teuk-concise}, written in index-free form as
\begin{align}
\mathcal{O}\mathcal{T}[h^{(1)}_{ab}]=\mathcal{S}\mathcal{E}[h^{(1)}_{ab}], 
\end{align}
%
is valid for any $h^{(1)}_{ab}$. Taking the adjoint of Eq.~\eqref{eq:Wald-OI} gives us another operator identity,
\begin{align}\label{eq:Wald-adjoint}
\mathcal{T}^\dagger\mathcal{O}^\dagger=\mathcal{E}\mathcal{S}^\dagger,
\end{align}
where we have used the fact that $\mathcal{E}$ is self-adjoint. It follows that if we can find a scalar field $\Phi^{(1)}$ satisfying ${\cal O}^\dagger\Phi^{(1)}=0$, then 
\beq
h^{(1)}_{ab}=2{\rm Re}(\mathcal{S}^\dagger_{ab}\Phi^{(1)})
\eeq
is a solution to the vacuum Einstein equation $\mathcal{E}_{ab}[h^{(1)}_{ab}]=0$. The structure of ${\cal S}$ in Eq.~\eqref{eq:S} immediately implies that the reconstructed metric perturbation is in a traceless outgoing radiation gauge (ORG), satisfying $h^{(1)}_{ab}n^b=0=h^{(1)}_{ab}g^{ab}$.\footnote{Despite being defined from the ingoing principal null vector, CCK reconstruction in this gauge yields a solution that is regular at \scri for outgoing radiation. Conversely, CCK reconstruction defined from the outgoing null vector $l^a$ yields a solution that is regular for ingoing radiation. See Ref.~\cite{Keidl:2010pm}.} The field $\Phi^{(1)}$, referred to as the ORG Hertz potential, can be obtained from $\psi^{(1)}_{4}$ using the circularity condition $\psi^{(1)}_4 = {\cal T}[h^{(1)}_{ab}] = 2{\cal T}[{\rm Re}(\mathcal{S}^\dagger\Phi^{(1)})]$, which enforces that the reconstructed perturbation $h^{(1)}_{ab}$ corresponds to the original Weyl scalar. This circularity condition can be reduced to a so-called radial inversion relation, a fourth-order ordinary differential equation along ingoing null rays, given in Eq.~\eqref{eq:Hertz}. In vacuum, the inversion relation can effectively be solved algebraically by simultaneously imposing ${\cal O}^\dagger\Phi^{(1)}=0$~\cite{ori2003reconstruction,van2015metric}.

More recently, GHZ extended the CCK method to nonvacuum perturbations ~\cite{green2020teukolsky}. In brief, they showed that the CCK procedure remains valid for any perturbation of Kerr with $T^{(1)}_{l\mu}=0$. They then provided a method for calculating a so-called \textit{corrector tensor}, $x^{(1)}_{ab}$, to account for the $T^{(1)}_{l\mu}\neq 0$ piece of the source. The total metric perturbation is hence\footnote{This can be modified by the addition of trivial perturbations toward another Kerr solution, which we choose to absorb into the background metric. Note, however, that $x^{(1)}_{ab}$ will include perturbations of that form in any vacuum region if there is a change in mass or angular momentum between two regions. For example, in the case of a point mass at radial position $r_p(t)$, $x^{(1)}_{ab}$ will include perturbations of the form 
\beq
\left(\delta M\frac{\partial g^{(0)}_{ab}}{\partial M}+\delta J \frac{\partial g^{(0)}_{ab}}{\partial J}\right)\theta(r-r_p), 
\eeq
where $\delta M$ and $\delta J$ are the particle's orbital energy and angular momentum, and $\theta$ is a step function~\cite{Toomani:2021jlo}. Due to the presence of the step function, these are not trivial vacuum perturbations and cannot be absorbed into the background metric.}
\begin{align}\label{eq:green-metric-recon}
h^{(1)}_{ab} = 2{\rm Re}(S^{\dagger}_{ab}\Phi^{(1)}) + x^{(1)}_{ab}.
\end{align}
$\Phi$ can still be found from $\psi^{(1)}_4$ by integrating the same radial inversion relation~\eqref{eq:Hertz}. If $x^{(1)}_{ab}$ is put in a traceful outgoing radiation gauge, meaning $x^{(1)}_{ab}n^b=0$ but $x^{(1)}_{m\mb}\neq0$,
then the field equation ${\cal E}_{ab}[x^{(1)}_{ab}]n^b = 8\pi T^{(1)}_{ab}n^b$ also reduces to a set of ordinary differential equations along ingoing null rays. Therefore, the complete metric perturbation can be found by solving the Teukolsky equation for $\psi^{(1)}_4$ and integrating the ordinary differential equations for $\Phi^{(1)}$ and $x^{(1)}_{ab}$. We refer to Appendix~\ref{app:GreenTeuk} for additional details.

These methods work in the same way if starting from $\psi^{(1)}_0$ and its associated operators ${\cal O}'$, ${\cal T}'$, and ${\cal S}'$ (using the GHP prime notation described in Appendix~\ref{app:GHP}). The resulting metric perturbation then satisfies the ingoing radiation gauge (IRG) condition $h^{(1)}_{ab}l^b=0$, and the piece obtained from the IRG Hertz potential is again traceless. The IRG Hertz potential can be obtained from $\psi^{(1)}_0$ by solving a fourth-order ordinary differential equation along outgoing null rays, given in Eq.~\eqref{eq:IRG Hertz}.\footnote{One can alternatively use ``angular inversion relations'' to obtain the ORG Hertz potential from $\psi^{(1)}_0$ and the IRG Hertz potential from $\psi^{(1)}_4$. See Table~I in Ref.~\cite{Keidl:2010pm} or Eqs.~(68) and~(71) in Ref.~\cite{Pound-Wardell:2021}.} Likewise, the field equation for the IRG corrector tensor, ${\cal E}_{ab}[x^{(1)}_{ab}]l^b = 8\pi T^{(1)}_{ab}l^b$, reduces to ordinary differential equations along outgoing null rays.

Other, similar nonvacuum reconstruction methods are also now becoming available, either in the Aksteiner--Andersson--B\"{a}ckdahl gauge~\cite{Aksteiner:2016pjt,Wardell:Capra2022,coupling} or in the Lorenz gauge~\cite{Dolan:2021ijg,Toomani:Lorenz,Dolan:Capra2022}. The emergence of all these methods partially motivates our choice of preferred second-order formalism in this paper.

\subsection{\label{sec:Self-forcetheory}Self-force theory}

At least through second order in perturbation theory, GSF theory can be reduced to solving the perturbative Einstein equations~\eqref{eq:linearEFE} and \eqref{eq:quadraticEFE} with $T^{(1)}_{ab}$ and $T^{(2)}_{ab}$ extracted from the first- and second-order terms in the expansion of the Detweiler stress-energy tensor~\cite{detweiler2012second-order,Upton},
\begin{align}
T_{ab} = m\int_\gamma \tilde u_a\tilde u_b\tilde\delta^4(x,x_p(\tilde\tau))d\tilde\tau.\label{eq:Detweiler T}
\end{align}
This represents a point mass $m$ moving on a worldline $\gamma$ (with coordinates $x_p^\alpha$) in a certain \emph{effective} vacuum spacetime
\beq
\tilde g_{ab} = g^{(0)}_{ab}+\varepsilon h^{{\rm R}(1)}_{ab}+\varepsilon^2 h^{{\rm R}(2)}_{ab}+\ldots,
\eeq
where $h^{{\rm R}(n)}_{ab}$ is a certain regular (smooth) piece of $h^{(n)}_{ab}$. In $T_{ab}$,  $\tilde\tau$ is the proper time in that metric, the four-velocity has components $\tilde u_\alpha=\tilde g_{\alpha\beta}dx_p^\beta/d\tilde\tau$, and the covariant delta function written in coordinate form is  
\beq
\tilde\delta^4(x,x_p(\tilde\tau))=\frac{\delta^4(x^\alpha-x^\alpha_p(\tilde\tau))}{\sqrt{-{\rm det}(\tilde g_{\mu\nu})}}. 
\eeq
The particle's trajectory $\gamma$ satisfies the geodesic equation in the effective metric,
\begin{align}
\tilde u^b\tilde\nabla_b\tilde u^a = O(\varepsilon^3).\label{EOM}
\end{align}
When written in terms of background proper time $\tau$ and the background derivative $\nabla_a$, this becomes~\cite{pound2015gaugeandmotion}
\begin{align}
\!\!\!\!u^b\nabla_b u^a &= -\frac{1}{2}P^{ab}\left(g_b{}^c-h^{{\rm R}\, c}_{b}\right)\left(2h^{\rm R}_{c(d;e)}-h^{\rm R}_{de;c}\right)u^d u^e\notag \\
 & \ \ \ +O(\varepsilon^3),\label{EOM v2}
\end{align}
where $u^\alpha=dx_p^\alpha / d\tau$, $P^{ab}:=(g^{ab}+u^a u^b)$, and $h^{{\rm R}}_{ab}:=\varepsilon h^{{\rm R}(1)}_{ab} +\varepsilon^2 h^{{\rm R}(2)}_{ab}$. 

At first order, most commonly, one solves directly for the retarded field $h^{(1)}_{ab}$ with the point-mass source $T^{(1)}_{ab}$ and then extracts $h^{{\rm R}(1)}_{ab}$ through subtraction of an appropriate singular piece~\cite{barack2009GSF-Implementation-Review,barackpound18}. At second order, no analogous calculation has been attempted due to the extreme singularities that appear  on $\gamma$ in the source term $\delta^2G_{ab}[h^{(1)}_{ab},h^{(1)}_{ab}]$. Because of these, a mathematically meaningful form of the field equation for the physical field $h^{(2)}_{ab}$, valid on the entire domain including $\gamma$, was only recently derived~\cite{Upton}, and it has not yet been cast in a practical form for computations. Instead, second-order GSF theory has generally been formulated, and always been implemented, using a puncture scheme. This scheme (like the description in terms of a point mass in $\tilde g_{ab}$) is derived from the method of matched asymptotic expansions. It ensures that the solution at small distances from $\gamma$ matches the metric outside a small compact body.

In a puncture scheme, one splits the physical field into two pieces, ${h}^{(n)}_{ab} = {h}^{\P(n)}_{ab} + {h}^{\res(n)}_{ab}$. The first piece, $h^{\P(n)}_{ab}$, is an analytically known ``puncture''~\cite{Pound-Miller:14} that encodes the dominant behavior of the local field outside the small object. It generically diverges as 
\beq\label{eq:hPn behavior}
h^{\P(n)}_{ab}\sim 1/\varrho^n 
\eeq
on $\gamma$, where $\varrho$ is proper spatial distance from $\gamma$. The puncture is confined to a finite region $\Gamma$ around the worldline $\gamma$, going to zero outside that region. The residual field $h^{\res(n)}_{ab}={h}^{(n)}_{ab}-{h}^{\P(n)}_{ab}$ is regular on $\gamma$; furthermore, for an appropriate puncture, $h^{\res(n)}_{ab}$ and $h^{\res(n)}_{ab;c}$ locally reduce to $h^{{\rm R}(n)}_{ab}$ and $h^{{\rm R}(n)}_{ab;c}$ on $\gamma$, such that they can be used in Eq.~\eqref{EOM v2}. Outside of $\Gamma$, $h^{\res(n)}_{ab}$ is equal to the physical field $h^{(n)}_{ab}$. 

After making this split, we rewrite the Einstein equations as equations for $h^{\res(n)}_{ab}$. Beginning from the vacuum field equations off the worldline, we move the punctures to the right-hand side and treat their contributions as effective sources for the residual fields:
\begin{align}
\delta G_{ab}[{{h}}^{\res(1)}_{ab}] &= -\delta G_{ab}[{{h}}^{\P(1)}_{ab}]\label{EFE1 - eff},\\
\delta G_{ab}[{{h}}^{\res(2)}_{ab}] &= - \delta^2 G_{ab}[{h}^{(1)}_{ab},{h}^{(1)}_{ab}] -\delta G_{ab}[{{h}}^{\P(2)}_{ab}].\label{EFE2 - eff}
\end{align}
These equations are valid on the entire domain, including $\gamma$, if we evaluate the sources as ordinary functions off $\gamma$ and then take the limit to $\gamma$; see Sec.~I of Ref.~\cite{Upton}. Individually, each of the two source terms in Eq.~\eqref{EFE2 - eff}  diverges as $1/\varrho^4$ on $\gamma$: $\delta G_{ab}[h^{\P(2)}_{ab}]\sim \nabla_c\nabla_d \varrho^{-2}$ and $\delta^2G_{ab}[{h}^{(1)}_{ab},{h}^{(1)}_{ab}]\sim (\nabla_c \varrho^{-1})^2$. But their sum is integrable there. The retarded solutions to these equations, when added to the punctures $h^{\P(n)}_{ab}$, are guaranteed to satisfy the original physical problem.

Nearly all first-order GSF calculations in Kerr have been performed in a so-called ``no-string'' radiation gauge~\cite{pound2014radgauge}, in which $\psi^{(1)}_0$ or $\psi^{(1)}_4$ is first calculated,  CCK reconstruction is applied in the vacuum region on either side of the particle's orbit, and the metric is then ``completed'' by the addition of appropriate mass, spin, and gauge perturbations in each of the two vacuum regions~\cite{Shah:2012gu,van2015metric,merlinpound2016,van2017mass,Shah:2015nva}. In this gauge, $h^{(1)}_{ab}$ includes delta functions on the sphere $r=r_p(t)$ that intersects the particle, along with jump discontinuities across it.\footnote{This contrasts with ``half-string'' and ``full-string'' radiation gauges, in which a stringlike singularity extends from the particle to infinity and/or to the horizon~\cite{pound2014radgauge}. See Ref.~\cite{Toomani:2021jlo} for a thorough analysis of this stringlike structure.} It is not known how to use such a singular metric perturbation as input in the second-order source. Two of us, with collaborators, recently showed how this problem can be overcome using GHZ reconstruction within a puncture scheme~\cite{Toomani:2021jlo,Bourg:GHZ} to calculate a sufficiently regular $h^{(1)}_{ab}$. The method solves Eq.~\eqref{EFE1 - eff} by applying the GHZ scheme to reconstruct the residual field $h^{\res(1)}_{ab}$ in a radiation gauge; adding the puncture $h^{\P(1)}_{ab}$ in any well-behaved gauge then yields the total $h^{(1)}_{ab}$. Work is also ongoing to instead reconstruct $h^{(1)}_{ab}$ in the Lorenz gauge~\cite{Dolan:2021ijg,Dolan:Capra2022,Toomani:Lorenz}, which is free from the radiation gauge's pathologies. 

Although GSF calculations at second order have relied on puncture schemes, and there is a push in that direction even at first order, our view is that there are likely to be continued advantages to directly solving for the physical field in some cases. For example, if one is calculating asymptotic fluxes, one does not require the regular field at the particle.

\subsection{Fluxes and adiabatic evolution}\label{subsec:balance-laws}

The most advanced first-order GSF codes calculate both the dissipative and conservative piece of the first-order GSF from a ``no-string'' metric perturbation $h^{(1)}_{ab}$~\cite{van2018gravitational}. But it is possible to calculate the dominant effects of the first-order GSF directly from $\psi^{(1)}_4$. This is possible because the dominant, 0PA phase, $\varphi^{(0)}$ in Eq.~\eqref{eq:frequencies}, can be obtained from the radiative ``half-retarded minus half-advanced'' field $h^{{\rm Rad}(1)}_{ab}=\frac{1}{2}\left(h^{{\rm Ret}(1)}_{ab}-h^{{\rm Adv}(1)}_{ab}\right)$~\cite{Mino:03}. Because it is the difference between two particular solutions, $h^{{\rm Rad}(1)}_{ab}$ is a vacuum perturbation. This means CCK reconstruction can be straightforwardly applied to evaluate the dissipative GSF directly in terms of modes of $\psi^{{\rm Rad}(1)}_4$, which are in turn readily expressed in terms of modes of the physical, retarded $\psi^{(1)}_{4}$~\cite{Sago-etal:05,Isoyama:2018sib}. 

Evolving the orbit in this way is closely related to balance-law arguments. For equatorial orbits, the 0PA orbital evolution is specified by the rates of change of orbital energy $E$ and angular momentum $L$. These rates of change are equal to the flux of energy and angular momentum down the horizon and out to infinity~\cite{Galtsov:82}, and the formulas for $dE/dt$ and $dL/dt$ obtained from the local GSF are identical to these fluxes, expressed in terms of modes of $\psi^{(1)}_{4}$. However, for inclined orbits, one must also track the evolution of the Carter constant $Q$. There is no known way to calculate this without directly substituting $h^{{\rm Rad}(1)}_{ab}$ into the formula for the local GSF (even if the resulting formula for $dQ/dt$ is often dubbed a ``flux-balance law'' in the literature). These points will be important in the context of post-adiabatic evolution. We defer further discussion to Sec.~\ref{sec:conclusion}, after we have introduced methods of calculating $\psi^{(2)}_4$.

\section{\label{sec:second-order-scheme I}Campanelli and Lousto's second-order Teukolsky equation}


Campanelli and Lousto~\cite{l-c} were the first to extend the Teukolsky equation to second and higher order. We now review their formulation and discuss its potential applications in second-order GSF calculations.

In this and the next section, we only give formulas for $\psi^{(2)}_4$. Formulas for $\psi^{(2)}_0$ can be obtained from the ones for $\psi^{(2)}_4$ by applying the GHP prime operation.



\subsection{Overview}\label{sec:LC-Equation}

Campanelli and Lousto's higher-order extension of the Teukolsky equation can be obtained following the same steps as at first order: projecting the Penrose wave equation~\eqref{eq:Penrose wave equation} onto appropriate tetrad legs and expanding all quantities in powers of $\varepsilon$. Alternatively, as Teukolsky did at first order, one can derive it from the Ricci and Bianchi identities in NP form; this was the route taken by Campanelli and Lousto. 
Either approach, when the expansions are carried to second order, produces a second-order Teukolsky equation,
\begin{align}\label{eq:LCT}
\mathcal{O}[\psi^{(2)}_4] &= 8\pi\mathcal{S}^{(2)}\Bigl[T^{(1)}_{ab},T^{(2)}_{ab},h^{(1)}_{ab},e^{(1)a}_{[\mu]}\Bigr] \notag \\
& \ \ \ + S^{(2)}_{CL}\Bigl[h^{(1)}_{ab},e^{(1)a}_{[\mu]}\Bigr],
\end{align}
with nonlinear source terms
\begin{align}\label{eq:LCT-Source}
S&^{(2)}_{CL}\Bigl[h^{(1)}_{ab},e^{(1)a}_{[\mu]}\Bigr] \notag\\
 &= \Bigl[ \bar{d}_3^{(0)} (\bm{\delta} +4\beta-\tau)^{(1)} - \bar{d}_4^{(0)}(\bm{D}+4\epsilon-\rho)^{(1)}\Bigr]\psi_4^{(1)} \notag \\
&\quad -\Bigl[\bar{d}_3^{(0)} (\bm{\Delta} +4\mu +2\gamma)^{(1)} - \bar{d}_4^{(0)}(\bar{\bm{\delta}}+4\pi+2\alpha)^{(1)}\Bigr]\psi_3^{(1)}\notag\\
&\quad +3\Big[ \bar{d}_3^{(0)} \nu^{(1)} - \bar{d}_4^{(0)}\lambda^{(1)}\Big]\psi_2^{(1)}\notag\\
&\quad+ 3 \Big[(\bar{d}_3-3\pi)^{(1)}\nu^{(1)}-(\bar{d}_4 - 3\mu)^{(1)}\lambda^{(1)}\Big]
\end{align}
and 
\begin{align}\label{eq:LCT2MatterSource}
\mathcal{S}&^{(2)}\Bigl[T^{(1)}_{ab},T^{(2)}_{ab},h^{(1)}_{ab},e^{(1)a}_{[\mu]}\Bigr] \notag\\
&= \frac{1}{2} \sum_{p=1}^{2} \bigg\{\bar{d}_4^{(0)} \Big[ (\bar{\bm{\delta}}-2\bar{\tau}+2\alpha)^{(2-p)}T^{(p)}_{n \bar{m}}  \notag \\
&\qquad\qquad\qquad- (\bm{\Delta}+2\gamma-2\bar{\gamma}+\bar{\mu})^{(2-p)}T^{(p)}_{\bar{m} \bar{m}} \Big] \notag \\ 
&\quad\qquad +  \bar{d}_3^{(0)} \Big[ (\bm{\Delta}+2\gamma+2\bar{\mu})^{(2-p)}T^{(p)}_{n \bar{m}} \notag \\
&\qquad\qquad\qquad - (\bar{\bm{\delta}}-\bar{\tau}+2\bar{\beta}+2\alpha)^{(2-p)}T^{(p)}_{n n} \Big] \bigg\}. 
\end{align}
These quantities involve the perturbed spin coefficients ($\kappa^{(1)}, \ \sigma^{(1)},\ldots$), which depend on both $h^{(1)}_{ab}$ and on $e^{(1)a}_{[\mu]}$. In Appendix~\ref{app:1st-order-NP} we express $e^{(1)a}_{[\mu]}$ and the spin coefficients entirely in terms of $h^{(1)}_{ab}$ (with some corrections to the analogous expressions that appeared in Ref.~\cite{l-c}). However, we note that doing so is only possible by specifying a choice of perturbed tetrad; generically, $e^{(1)a}_{[\mu]}$ contains infinitesimal rotation freedom that cannot be specified by $h^{(1)}_{ab}$. 

To understand how $\mathcal{S}^{(2)}$ relates to the source operator $\mathcal{S}$ in the first-order Teukolsky equation, one can split 
Eq.~\eqref{eq:LCT2MatterSource} into two pieces,
\begin{multline}\label{eq:LCTMatterSourceSimple}
\mathcal{S}^{(2)}\Bigl[T^{(1)}_{ab},T^{(2)}_{ab},h^{(1)}_{ab},e^{(1)a}_{[\mu]}\Bigr] \\= \mathcal{S}[T^{(2)}_{ab}]+\mathcal{S}_{*}\Bigl[T^{(1)}_{ab},h^{(1)}_{ab},e^{(1)a}_{[\mu]}\Bigr],
\end{multline}
where
\begin{align}\label{eq:LCTprimeMatterSource}
\mathcal{S}&_{*}\Bigl[T^{(1)}_{ab}, h^{(1)}_{ab},e^{(1)a}_{[\mu]}\Bigr]\notag\\
&:= \frac{1}{2} \bigg\{ \bar{d}_4^{(0)} \Big[ (\bar{\bm{\delta}}-2\bar{\tau}+2\alpha)^{(1)}T^{(1)}_{n \bar{m}}  \notag \\
&\qquad\qquad\quad - (\bm{\Delta}+2\gamma-2\bar{\gamma}+\bar{\mu})^{(1)}T^{(1)}_{\bar{m} \bar{m}} \Big] \notag \\ 
&\qquad  +  \bar{d}_3^{(0)} \Big[ (\bm{\Delta}+2\gamma+2\bar{\mu})^{(1)}T^{(1)}_{n \bar{m}} \notag \\
&\qquad\qquad\quad -  (\bar{\bm{\delta}}-\bar{\tau}+2\bar{\beta}+2\alpha)^{(1)}T^{(1)}_{n n} \Big] \bigg\}.
\end{align}
Comparing to Eq.~\eqref{eq:S}, we see that $\mathcal{S}_{*}\Bigl[\,\cdot\,,h^{(1)}_{ab},e^{(1)a}_{[\mu]}\Bigr]$, treated as a linear operator that acts on a stress-energy tensor, is a first-order correction to $\mathcal{S}$. The total source in Eq.~\eqref{eq:LCT} then reads
\beq\label{eq:LC total source}
8\pi\left(\mathcal{S}[T^{(2)}_{ab}] +\mathcal{S}_{*}\Bigl[T^{(1)}_{ab},h^{(1)}_{ab},e^{(1)a}_{[\mu]}\Bigr]\right)+S^{(2)}_{CL}\Bigl[h^{(1)}_{ab},e^{(1)a}_{[\mu]}\Bigr].
\eeq

Equation~\eqref{eq:LCT} has the same structure as the first-order Teukolsky equation~\eqref{eq:teuk-concise}. They only differ in having a different source. So, in particular, the second-order equation is separable in precisely the same way as at first order. In an appropriate gauge, $\psi^{(2)}_4$ also directly represents the second-order term in the asymptotic waveform, and from it one can compute asymptotic fluxes of energy and angular momentum using standard formulas; see, for example, Eqs.~(22) and (24) in Ref.~\cite{l-c}. However, the condition ``in an appropriate gauge'' is indispensable (and subtle) here, as we discuss in Sec.~\ref{sec:asymptotics}.

\subsection{Infinitesimal tetrad-rotation and gauge dependence of \texorpdfstring{$\psi_4^{(2)}$}{Lg} }\label{subsec:LCT-transform}

As pointed out by Campanelli and Lousto, $\psi_4^{(2)}$ (unlike $\psi_4^{(1)}$) is not invariant under infinitesimal tetrad rotations. This can be seen straightforwardly by applying an infinitesimal boost, as defined from Eq.~\eqref{eq:boost} with $A=1+O(\varepsilon)$. From the definition of $\psi_4$ in Eq.~\eqref{eq:psi4 def}, we have that the exact $\psi_4$ transforms to $A^{-2}\psi_4$. Expanding $A$ and $\psi_4$ in powers of $\varepsilon$, we see that $\psi_4^{(2)}$ transforms as
\beq
\psi^{(2)}_4 \to \psi^{(2)}_4 -2 A^{(1)}\psi^{(1)}_4,
\eeq
where we have used $\psi_4^{(0)}=0$ (we add a superscript zero here to avoid ambiguity). Since $\psi^{(1)}_4$ is nonzero except in trivial cases, $\psi_4^{(2)}$ is not invariant.
%
Campanelli and Lousto give a method for constructing an infinitesimal-tetrad-rotation-invariant quantity from $\psi_4^{(2)}$ by adding a quadratic combination of first-order terms.

Similarly, unlike $\psi_4^{(1)}$, $\psi_4^{(2)}$ is not gauge invariant. From Eq.~\eqref{DeltaA2}, $\psi_4^{(2)}$ transforms as
\begin{align}
\Delta\psi_4^{(2)} 
=\mathcal{L}_{\xi^c_{(1)}}\psi_4^{(1)},\label{eq:Delta psi42}
\end{align}
again using $\psi_4^{(0)}=0$.
Campanelli and Lousto also proposed a method for constructing a gauge-invariant quantity from $\psi_4^{(2)}$ by, effectively, transforming it to an ORG.

\subsection{Utility in GSF calculations\label{subsec:LCTsourceUnIterable}}

If we start from the second-order Einstein equation~\eqref{eq:quadraticEFE} with the source extracted from~\eqref{eq:Detweiler T}, it is not hard to see that the second-order Teukolsky equation~\eqref{eq:LCT} becomes ill-defined in the GSF context. 
The singular nature of $h^{(1)}_{ab}$ and $T^{(1)}_{ab}$ at the particle's worldline, $\gamma$, causes the source to be ill defined.

This is most striking in the source term $S_*[T^{(1)}_{ab},h^{(1)}_{ab}]$ defined in  Eq.~\eqref{eq:LCTprimeMatterSource}. It is composed of products of $T^{(1)}_{ab}$ and $h^{(1)}_{ab}$. Recall that $T^{(1)}_{ab}$ is a delta function supported on the particle and that $h^{(1)}_{ab} \sim \frac{1}{\varrho}$, where we recall that $\varrho$ is the proper spatial distance from $\gamma$. Their product therefore has the manifestly ill-defined form $T^{(1)}_{ab}h^{(1)}_{ab}\sim\frac{\delta(\varrho)}{\varrho}$. 

The other source terms are also problematic. As an example, we show that $S^{(2)}_{CL}[h^{(1)}_{ab},h^{(1)}_{ab}]$ is not locally integrable at the worldline. This source, defined in Eq.~\eqref{eq:LCT-Source}, is a complicated fourth-order differential operator, quadratic in $h^{(1)}_{ab}$. Hence, at the worldline, one can expect it to diverge as
\begin{align}
S^{(2)}_{CL}[h^{(1)}_{ab},h^{(1)}_{ab}]& \sim (\partial_\varrho \partial_\varrho h^{(1)}_{ab})(\partial_\varrho \partial_\varrho h^{(1)}_{ab}) \notag \\
&\sim \varrho^{-6}.
\end{align}
Integrating over a small region $\varrho<R$, we see
\begin{align}\label{eq:LCT-source-non-intergrable}
\int_0^R S^{(2)}_{CL}[h^{(1)}_{ab},h^{(1)}_{ab}] \varrho^2 d \Omega &\sim \int_0^R \varrho^{-6} \varrho^2 d\Omega\notag\\ 
&= \bigg[-\frac{4\pi}{3\varrho^3} \bigg]^R_0.
\end{align}
Clearly this diverges at the lower limit, meaning $S^{(2)}_{CL}[h^{(1)}_{ab},h^{(1)}_{ab}]$ is not locally integrable at $\varrho=0$.

One can avoid this problem of an ill-defined source by implementing a puncture scheme. As described in Sec.~\ref{sec:Self-forcetheory}, we first consider the field equation at all points off the worldline. This region is vacuum, with $T^{(1)}_{ab}=0=T^{(2)}_{ab}$, such that $\mathcal{O}[\psi^{(2)}_4] = S^{(2)}_{CL}[h^{(1)}_{ab},h^{(1)}_{ab}]$. To isolate the singular piece of $\psi_4^{(2)}$, we first fix the infinitesimal tetrad freedom as in Appendix~\ref{app:1st-order-NP}. This allows us to express $e^{(1)a}_{[\mu]}$ uniquely in terms of $h^{(1)}_{ab}$. The quadratic terms in the Weyl scalar, $\psi^{(2)}_{4Q}$ in Eq.~\eqref{eq:psi42deconstructed}, then becomes $\psi^{(2)}_{4Q}=\delta^2\psi_4[h^{(1)}_{ab},h^{(1)}_{ab}]$, where $\delta^2\psi_4$ is now bilinear in its arguments. We then split the metric perturbations into punctures and residual fields; i.e.,  $h^{(n)}_{ab}=h^{\res(n)}_{ab}+h^{\P(n)}_{ab}$. 
Now we can define $\psi^{\P(2)}_4$:
\begin{align}\label{eq:psi42deconstructed-puncture}
\psi_4^{\P(2)}&=\mathcal{T}[h^{\P(2)}_{ab}] + \delta^2\psi_4[h^{\P(1)}_{ab},h^{\P(1)}_{ab}] \notag \\
& \quad 
+\delta^2\psi_4[h^{\P(1)}_{ab},h^{\res(1)}_{ab}]  +\delta^2\psi_4[h^{\res(1)}_{ab},h^{\P(1)}_{ab}]   \notag \\
&=:\psi_{4L}^{\P(2)}+\psi_{4Q}^{\P(2)}.
\end{align}
Moving this puncture to the right-hand side of the second-order vacuum equation, we obtain a field equation for the residual $\psi^{\res(2)}_4=\psi^{(2)}_4-\psi^{\P(2)}_4$,
\begin{align}\label{eq:LCTpunc}
&\mathcal{O}[\psi^{\res(2)}_4] = S^{(2)}_{CL}[h^{(1)}_{ab},h^{(1)}_{ab}]-\mathcal{O}[\psi^{\P(2)}_4].
\end{align}
This can be extended down to the worldline to apply over the full spacetime.

In practice, the puncture $\psi^{\P(2)}_4$ would be obtained as an expansion in powers of $\varrho$~\cite{Pound-Miller:14}. The requirements on this expansion are quite severe. Each of the two source terms in Eq.~\eqref{eq:LCTpunc} generically diverges as $\varrho^{-6}$. Making the effective source integrable therefore requires cancelling four powers of $\varrho$. Given that $h^{\P(2)}_{ab}\sim \varrho^{-2}$ and that $\psi^{\P(2)}_4\sim \nabla\nabla h^{\P(2)}_{ab}\sim \varrho^{-4}$, the first four orders in $\psi^{\P(2)}_4$ have the following form:
\beq\label{eq:psi42P}
\psi^{\P(2)}_4 \sim \frac{1}{\varrho^4} + \frac{1}{\varrho^3} + \frac{1}{\varrho^2} + \frac{1}{\varrho}.
\eeq
Constructing this puncture would be possible using the highest-order expressions available for $h^{\P(2)}_{ab}$~\cite{Pound-Miller:14}. However, this will only suffice to make the source integrable. It will not suffice to make $\psi^{\res(2)}_4$ continuous at the particle; the $\varrho^0$ term in $\psi^{(2)}_4$ will generically have a directional discontinuity at the particle, or even a logarithmic divergence, and the $\psi^{\P(2)}_4$ in Eq.~\eqref{eq:psi42P} is not sufficiently high order to remove those discontinuities. It is not obvious whether metric reconstruction will then yield a differentiable residual field $h^{\res(2)}_{ab}$ (as would be required to use it in the GSF). We expect forthcoming work to alleviate this problem by pushing analytical punctures to a much higher order~\cite{Bourg:puncture}.

As an alternative to a puncture scheme, one can try to construct a well-defined distributional source for the physical field $\psi^{(2)}_4$. The problem of an ill-defined source also arose historically for the second-order Einstein equation, and Ref.~\cite{Upton} derived two ways around it. Unfortunately, neither approach can be straightforwardly applied to the source for $\psi^{(2)}_4$. 

One approach in Ref.~\cite{Upton}, building on the earlier~\cite{pound2017HighlyRegularGauge}, was to adopt a so-called ``highly regular gauge'', in which $h^{(2)}_{ab}\sim \varrho^{-1}$ rather than the generic $\varrho^{-2}$ suggested by Eq.~\eqref{eq:hPn behavior}. In that special gauge, the worst part of the source, $\ddG_{ab}[h^{\P(1)}_{ab},h^{\P(1)}_{ab}]$, becomes integrable, behaving as $\sim\varrho^{-2}$ rather than the generic $\sim\varrho^{-4}$. It is unlikely that this method will help for the Teukolsky source $S_{CL}$. Because $S_{CL}$ is two orders more singular than $\ddG_{ab}$, we can expect it to behave as $\sim\varrho^{-4}$ in a highly regular gauge, making it nonintegrable. 

The second method in Ref.~\cite{Upton} was to adopt a canonical distributional definition of $\ddG_{ab}[h^{(1)}_{ab},h^{(1)}_{ab}]$. This approach fails here because of the fundamental nonlinearity of $\psi^{(2)}_4$. In the case of the Einstein equation, the method works because even though $\delta^2G_{ab}[h^{(1)}_{ab},h^{(1)}_{ab}]$ is not locally integrable, we know that it is equal to $-\delta G_{ab}[h^{(2)}_{ab}]$ at each point away from the worldline. $h^{(2)}_{ab}$ \emph{is} integrable even in a generic gauge in which it diverges as $\varrho^{-2}$. Therefore $\delta G_{ab}[h^{(2)}_{ab}]$ is well defined as a distribution because it is a linear operator acting on an integrable function~\cite{friedlander1998distributiontheory}. Ref.~\cite{Upton} used this fact to promote $\delta^2G_{ab}[h^{(1)}_{ab},h^{(1)}_{ab}]$ to a well-defined distribution by effectively replacing its worst-behaving piece with the most singular piece of $-\delta G_{ab}[h^{(2)}_{ab}]$; see Sec. VA of Ref.~\cite{Upton}. We cannot follow an analogous approach here because ${\cal O}[\psi^{(2)}_{4}]$ is not a linear operator on an integrable function. Using the split~\eqref{eq:psi42deconstructed} of $\psi^{(2)}_{4}$ into its linear and quadratic pieces, we can divide ${\cal O}[\psi^{(2)}_{4}]$ into ${\cal O}[{\cal T}[h^{(2)}_{ab}]]$ plus ${\cal O}[\psi^{(2)}_{4Q}]$. The first term is a linear operator on the integrable function $h^{(2)}_{ab}$, but the second term is a linear operator on the nonintegrable, nonlinear function $\psi^{(2)}_{4Q}$.

One caveat to this analysis is that we have not accounted for the source's dependence on $e^{(1)a}_{[\mu]}$. It might be possible to use the freedom of infinitesimal tetrad rotations to improve the behavior of the source, just as it was possible to use the highly regular gauge to improve the behavior of the source in the second-order Einstein equation.

\section{\label{sec:pure-2nd-Teuk} Reduced second-order Teukolsky equation}

We now turn to the alternative formulation of the second-order Teukolsky equation. This equation has the same operator on the left-hand side as Eq.~\eqref{eq:LCT}, but it has a different field variable and a different source. We discuss its application to GSF calculations and how it integrates into emerging metric reconstruction methods.

\subsection{Overview}

As pointed out by Campanelli and Lousto~\cite{l-c}, the second-order Weyl scalar naturally splits into a piece ($\psi^{(2)}_{4L}$) that is linear in $h^{(2)}_{ab}$  and a piece ($\psi^{(2)}_{4Q}$) that is quadratic in the first-order quantities $h^{(1)}_{ab}$ and $e^{(1)a}_{[\mu]}$; refer back to Eq.~\eqref{eq:psi42deconstructed}. Campanelli and Lousto's second-order field equation was for the sum $\psi^{(2)}_{4L}+\psi^{(2)}_{4Q}$. Here we advocate for using a field equation for $\psi^{(2)}_{4L}$ alone. 

The derivation of such a field equation follows immediately from Wald's operator identity~\eqref{eq:Wald-OI}. Applying that identity to $h^{(2)}_{ab}$ produces a second-order Teukolsky equation for $\psi^{(2)}_{4L}$:
\begin{align}
\mathcal{O}\mathcal{T}[h^{(2)}_{ab}]&=\mathcal{S}\mathcal{E}[h^{(2)}_{ab}], \notag \\
\Rightarrow \mathcal{O}[\psi_{4L}^{(2)}]&=\mathcal{S}\big[8\pi T^{(2)}_{ab} - \delta^2 G_{ab}[h^{(1)}_{ab},h^{(1)}_{ab}]\big],\label{eq:pure-2nd-Teuk}
\end{align}
where we have used $\psi_{4L}^{(2)}=\mathcal{T}[h^{(2)}_{ab}]$ [from Eq.~\eqref{eq:psi42deconstructed}] and the second-order Einstein equation~\eqref{eq:quadraticEFE}. For want of a better name, we call Eq.~\eqref{eq:pure-2nd-Teuk} the reduced second-order Teukolsky equation.

This reduced equation appeared previously~\cite{green2020teukolsky,Spiers:GR22}, and in particular, it was a motivating factor in GHZ's development of their nonvacuum metric reconstruction scheme. GHZ reconstruction, as outlined in Sec.~\ref{sec:reconstruction}, is able to obtain a metric perturbation $h_{ab}$ satisfying $\delta G_{ab}[h_{ab}]=S_{ab}$, for any distributionally well-defined $S_{ab}$, starting from a solution to $\mathcal{O}[\psi]={\cal S}[S_{ab}]$. Therefore it can be used to reconstruct $h^{(2)}_{ab}$ from $\psi^{(2)}_{4L}$. The same is true for other nonvacuum reconstruction methods under development~\cite{Aksteiner:2016pjt,Toomani:Lorenz,Wardell:Capra2022,Dolan:Capra2022}. This provides one of our primary reasons for preferring the reduced equation. Ultimately, to calculate the second-order GSF, we require the complete second-order metric perturbation, making metric reconstruction methods crucial.

Given that the reduced equation is already available in the literature, our primary new contribution here is explicit expressions for its nonlinear source term ${\cal S}[\delta^2G_{ab}[h^{(1)}_{ab},h^{(1)}_{ab}]]$ in NP and GHP notation. That expression can be found in the accompanying \pkg{Mathematica} notebook~\cite{2nd-order-notebook}. We also provide all of the tetrad components of $\delta^2G_{ab}[h^{(1)}_{ab},h^{(1)}_{ab}]$, which can be used to calculate the analogous source for the linear piece of $\psi^{(2)}_{0}$, $\psi^{(2)}_{0L}$.



If, for some reason, the complete $\psi_{4}^{(2)}$ is needed in an application, one can easily construct it from $\psi_{4L}^{(2)}$ and $h^{(1)}_{ab}$ using Eq.~\eqref{eq:psi42deconstructed}. In most contexts, this will be unnecessary because $\psi_{4L}^{(2)}$ and $\psi_{4}^{(2)}$ encode the same content from $h^{(2)}_{ab}$. In particular, $\psi_{4L}^{(2)}$ and $\psi_{4}^{(2)}$ encode the same second-order contributions to the waveform. This can be understood by considering that $h^{(1)}_{ab}$ decays like $r^{-1}$ in an asymptotically flat gauge, and $\psi_{4Q}^{(2)}$ is quadratic in $h^{(1)}_{ab}$, implying $\psi_{4Q}^{(2)}$ must fall off like $r^{-2}$; hence, for $r\rightarrow\infty$,
\beq
\psi_{4}^{(2)}=\psi_{4L}^{(2)}+\mathcal{O}(r^{-2}).
\eeq
We discuss the asymptotic behavior of $\psi_{4}^{(2)}$ and $\psi_{4L}^{(2)}$ in more detail in Sec.~\ref{sec:asymptotics}.



\subsection{Infinitesimal-tetrad-rotation invariance and gauge dependence of \texorpdfstring{$\psi_{4L}^{(2)}$}{Lg} }\label{subsec:psi4L2-transform}

Solving the reduced second-order Teukolsky equation for $\psi_{4L}^{(2)}$ offers the advantage that $\psi_{4L}^{(2)}$ is invariant under infinitesimal tetrad rotations, unlike $\psi_4^{(2)}$. This invariance is trivial because the operator $\T$, given in Eq.~\eqref{eq:T}, involves only zeroth-order quantities. All the infinitesimal tetrad dependence in $\psi_4^{(2)}$ can therefore be attributed to the dependence of $\psi_{4Q}^{(2)}$ on the tetrad perturbations $e^{(1)a}_{[\mu]}$.

Correspondingly, the source in the reduced second-order Teukolsky equation~\eqref{eq:pure-2nd-Teuk} is also trivially invariant under infinitesimal tetrad rotations because $\S$, given in Eq.~\eqref{eq:S}, likewise depends only on background quantities.

Still, $\psi_{4L}^{(2)}$, like $\psi_{4}^{(2)}$, is gauge dependent. Using the gauge-transformation rule~\eqref{Deltah2}, we find that $\psi_{4L}^{(2)}$ transforms as
\begin{align}
\Delta\psi_{4L}^{(2)}&=\mathcal{T}[\Delta h^{(2)}_{ab}] \notag\\
&=\mathcal{T}\left[\mathcal{L}_{\xi_{(1)}^c} h^{(1)}_{ab} + \tfrac{1}{2}\mathcal{L}_{\xi_{(1)}^c}\mathcal{L}_{\xi_{(1)}^c} g_{ab}\right].\label{eq:phi42GT}
\end{align}
Due to Eq.~\eqref{eq:Delta dpsi}, no second-order gauge vector appears in this transformation. That is, $\psi_{4L}^{(2)}$ is invariant under a purely second-order gauge transformation. But like $\psi^{(2)}_4$, it does transform under a change of first-order gauge due to the presence of $\xi^a_{(1)}$ in Eq.~\eqref{eq:phi42GT}. From Eqs.~\eqref{eq:phi42GT} and \eqref{eq:Delta psi42}, we can also deduce a relatively simple transformation rule for the quadratic part of $\psi^{(2)}_4$,
\beq
\Delta\psi^{(2)}_{4Q} = {\cal L}_{\xi^a_{(1)}}\psi^{(1)}_{4} - {\cal T}[\Delta h^{(2)}_{ab}].
\eeq

A reader might observe that the transformation~\eqref{eq:phi42GT} of $\psi^{(2)}_{4L}$ is substantially more complicated than the transformation of $\psi^{(2)}_{4}$. However, in practice, there would rarely be a need to invoke either of these rules. Once boundary conditions are specified, the gauge of $\psi^{(2)}_{4L}$ is implicitly determined by the gauge of the source term in the reduced second-order Teukolsky equation; the same is true of $\psi^{(2)}_4$ and its source. The gauge of the source, in turn, is determined entirely by the gauge of $h^{(1)}_{ab}$. In one of the follow-up papers~\cite{paperII}, we will use this fact to construct a gauge-independent version of $\psi_{4L}^{(2)}$ (and of its source).

\subsection{Utility in GSF calculations \label{sec:puncture}}

The important difference between Eq.~\eqref{eq:LCT} and Eq.~\eqref{eq:pure-2nd-Teuk} is the form of their respective sources. As discussed in Sec.~\ref{subsec:LCTsourceUnIterable}, due to the singular behavior of $h^{(1)}_{ab}$ and $T^{(1)}_{ab}$ on $\gamma$, the source in Eq.~\eqref{eq:LCT} requires special treatment in GSF calculations. Here we show the more desirable properties of the source in Eq.~\eqref{eq:pure-2nd-Teuk}.

We first simply observe that (i) the Teukolsky source operator $\mathcal{S}$ is linear, and (ii) it acts on a well-defined distribution $8\pi T^{(2)}_{ab} - \delta^2 G_{ab}[h^{(1)}_{ab},h^{(1)}_{ab}]$~\cite{Upton}. Therefore basic distribution theory tells us that $\mathcal{S}\big[8\pi T^{(2)}_{ab} - \delta^2 G_{ab}[h^{(1)}_{ab},h^{(1)}_{ab}]\big]$ is well defined as a distribution~\cite{friedlander1998distributiontheory}.



Formulatng a puncture scheme is equally straightforward. 
%
%
We simply apply the Wald identity~\eqref{eq:Wald-OI} to $h^{\res(2)}_{ab}$. Equation~\eqref{EFE2 - eff} then gives us
\begin{align} \label{eq:Linearised-2nd-Teuk-Puncture2}
 \mathcal{O}[\psi_{4L}^{\res(2)}]=-\mathcal{S}\big[ \delta^2G_{ab}[h^{(1)}_{ab},h^{(1)}_{ab}]\big]-\mathcal{O}[\psi_{4L}^{\P(2)}],
\end{align}        
where $\psi_{4L}^{\res(2)}:={\cal T}[h^{\res(2)}_{ab}]$ and $\psi_{4L}^{\P(2)}:={\cal T}[h^{\P(2)}_{ab}]$. The two source terms each diverge strongly, like $\sim\partial^2_{\varrho}\varrho^{-4}\sim \varrho^{-6}$ (or more mildly in a highly regular gauge). But they cancel to yield an integrable effective source. We refer back to Sec.~\ref{subsec:LCTsourceUnIterable} for a discussion of the requirements on the puncture; the analysis around Eq.~\eqref{eq:psi42P} also holds true for $\psi_{4L}^{\P(2)}$.



\section{Asymptotic behavior of the second-order Teukolsky equation}\label{sec:asymptotics}

In the preceding sections, we focused on the behavior of the second-order Teukolsky source on a particle's worldline. To complement that discussion, we now analyse the source in the opposite extreme: asymptotically, near \scri. Slow asymptotic falloff has been one of the main difficulties in second-order GSF calculations~\cite{PoundLargeScales}. Our conclusion in this section is that the problem persists in the second-order Teukolsky equation, but it can be eliminated by choosing a gauge that is adapted to the spacetime's null structure.

To analyse the source's falloff, we adopt retarded coordinates $(u,r,\theta,\phi)$ and the Kinnersley tetrad. At large $r$, the tetrad legs then behave as~\cite{Toomani:2021jlo}
\begin{align}
l^\mu\partial_\mu &= \partial_r,\\
n^\mu\partial_\mu &\sim \partial_u - \tfrac{1}{2}\partial_r + {\cal O}(r^{-1}), \\
m^\mu\partial_\mu&\sim \frac{1}{r}(ia\partial_u + \partial_\theta+\partial_\phi)+{\cal O}(r^{-2}).
\end{align}
Since all spin coefficients behave as $\order{r^{-1}}$~\cite{Pound-Wardell:2021}, the large-$r$ behavior of the tetrad legs immediately carries over to the directional derivatives $\bm{D}$, $\bm{\Delta}$, and $\bm{\delta}$. Hence, at large $r$, $\bm{\Delta}\sim\partial_u$ leaves the order in $r$ unchanged, while all other derivatives lower the order by one power in $r$. 

We assume the matter source $\S[T^{(2)}_{ab}]$ is spatially bounded, such that it does not contribute at large $r$. We next note that, from Eq.~\eqref{eq:S}, only $\ddG_{\mb\mb}$, $\ddG_{n\mb}$, and $\ddG_{nn}$ appear in the source $\S[\ddG_{ab}]$. Given the scalings from the previous paragraph, we see
\begin{align}\label{eq:GHPSfalloff}
\S[\ddG_{ab}]\sim \ddG_{\mb\mb} +\frac{\ddG_{n\mb}}{r} + \frac{\ddG_{nn}}{r^2}.
\end{align}
In words, the Teukolsky source has the same large-$r$ scaling as the source in the Einstein equation, but the behavior depends sensitively on the falloff of each component of the Einstein source. By analysing the form of $\delta^2 G_{ab}[h^{(1)}_{ab},h^{(1)}_{ab}]$, we will find the necessary gauge conditions on $h^{(1)}_{ab}$ to accelerate the falloff. 

Firstly, note the trace piece of $\ddG_{ab}$ does not appear in $\S[\ddG_{ab}]$, and the first-order perturbation is vacuum away from the worldline; therefore, analysing the falloff of $\delta^2 R_{ab}[h^{(1)}_{ab},h^{(1)}_{ab}]$ is sufficient. To determine that falloff, we analyse the large-$r$ behavior of $h^{(1)}_{\mu\nu}$, $h^{(1)}_{\mu\nu;\gamma}$ and $h^{(1)}_{\mu\nu;\gamma\delta}$. First, we assume that the metric perturbation is asymptotically flat at \scri, such that in a well-behaved gauge, we can write
\begin{align}\label{eq:h1 asymptotic}
    h^{(1)}_{\mu\nu} = \frac{z_{\mu\nu}(u,\theta,\phi)}{r} +\order{r^{-2}},
\end{align}
where $z_{\mu\nu}(u,\theta,\phi)$ is an arbitrary function of $u$, $\theta$, and $\phi$. This implies
\begin{align}
h^{(1)}_{\mu\nu;\gamma}&=-n_\gamma \bm{D}h^{(1)}_{\mu\nu}-l_\gamma\bm{\Delta} h^{(1)}_{\mu\nu} + \bar{m}_\gamma\bm{\delta} h^{(1)}_{\mu\nu} + m_\gamma\bar{\bm{\delta}}h^{(1)}_{\mu\nu} \notag \\
&= -\frac{\dot z_{\mu\nu}(u,\theta,\phi)}{r}l_\gamma +\order{r^{-2}},\label{eq:hsinglecovd}
\end{align}
where an overdot denotes differentiation with respect to $u$. Similarly, 
\begin{align}
h^{(1)}_{\mu\nu;\gamma\delta}
&= \frac{\ddot z_{\mu\nu}(u,\theta,\phi)}{r}l_\delta l_\gamma + \order{r^{-2}}. \label{eq:doublecovd}
\end{align}
Substituting these into the covariant formula~\eqref{eq:d2Rab} for $\ddR_{ab}$, we find the generic behavior 
\beq
\ddG_{\mu\nu} = \frac{F_{\mu\nu}(u,\theta,\phi)}{r^2}+\order{r^{-3}}
\eeq
for some $F_{\mu\nu}$ constructed from quadratic products of $z_{\mu\nu}$, $\dot z_{\mu\nu}$, and $\ddot z_{\mu\nu}$. Therefore, in a generic gauge, Eq.~\eqref{eq:GHPSfalloff} implies $\S[\ddG_{ab}[h^{(1)}_{ab},h^{(1)}_{ab}]]\sim r^{-2}$.

This is the same problematic falloff explored in Ref.~\cite{PoundLargeScales} (see also~\cite{Miller-Pound:2020}). It causes two problems for any field equation that asymptotically behaves like a flat-spacetime scalar wave equation, including the Teukolsky equation~\eqref{eq:teuk-concise}. The first problem is that for nonstationary source modes $S_\omega\sim e^{-i\omega u}/r^2$ with frequencies $\omega\neq0$, the solution falls off like $\sim \ln(\omega r)e^{-i\omega u}/(\omega r)$, rather than like a freely propagating wave $\sim e^{-i\omega u}/r$. This behavior is easily deduced from the form of the flat-spacetime scalar wave operator in retarded coordinates, $\sim r^{-1}(\partial_r^2 -2\partial_u\partial_r)(r\,\cdot)$. The second problem is that for stationary ($\omega=0$) source modes $S_{0}\sim r^{-2}$, the integral of the source against the retarded Green's function ($G_{\omega=0}\sim r^{-1}$) does not converge. It behaves as $\sim\int G_{0}S_{0} r^2drd\Omega\sim \int r^{-1}drd\Omega$, which diverges logarithmically as the upper limit of integration is taken to $r\to\infty$.

There are two reasons to expect that these ill behaviors should be a gauge artefact. According to the peeling theorem \cite{penrose1965zero}, the exact, fully nonlinear $\psi_{4}$ should fall off as $r^{-1}$, which should carry over to $\psi^{(2)}_{4L}$. Moreover, in vacuum at large $r$ it should behave as a freely propagating wave, meaning the $r^{-1}$ term should satisfy the homogeneous equation at leading order. This is only possible if the source for $\psi^{(2)}_{4L}$ decays as $r^{-4}$. 

Another reason to expect this $r^{-4}$ behavior is that $\delta^2 G_{ab}[h^{(1)}_{ab},h^{(1)}_{ab}]$ is effectively the stress-energy tensor of gravitational waves near \scri, which should have the form of outgoing null radiation. The piece carrying energy flux should behave like $\sim l_a l_b/r^2$, appearing as a $r^{-2}$ term in $\delta^2 G_{nn}[h^{(1)}_{ab},h^{(1)}_{ab}]$. The piece carrying angular momentum flux should behave like $\sim l_a m_b/r^3$, appearing as a $r^{-3}$ term in $G_{n\bar{m}}[h^{(1)}_{ab},h^{(1)}_{ab}]$. Neither piece has a component $\sim m_a m_b$, suggesting $G_{\bar{m}\bar{m}}[h^{(1)}_{ab},h^{(1)}_{ab}]$ should appear at $\order{r^{-4}}$. Hence, we expect that, in an appropriate gauge,  
\begin{align}
\delta^2 G_{nn} \sim r^{-2}, \quad \delta^2 G_{n\bar{m}}& \sim r^{-3}, \notag\\ \delta^2 G_{\bar{m}\bar{m}} \sim r^{-4}&. \label{d2Gfalloffs}
\end{align}
Referring to Eq.~\eqref{eq:GHPSfalloff}, we see that this would imply $\S[\ddG_{ab}]\sim r^{-4}$. 

We find it is possible to enforce the behavior~\eqref{d2Gfalloffs} by adopting a Bondi--Sachs gauge~\cite{madler2016bondisachs}. In such a gauge, the metric perturbation satisfies the gauge conditions $h^{(1)}_{\mu\nu}\sim r^{-1}$, $h^{(1)}_{ln}\sim h^{(1)}_{lm}\sim h^{(1)}_{m\mb} \sim r^{-2}$, and $h^{(1)}_{ll} \sim r^{-3}$. These conditions are adapted to the spacetime's lightcone structure by enforcing that, asymptotically, (i) the outgoing null vector $l^a$ remains null in the perturbed spacetime, and (ii) spheres of constant $(u,r)$ have surface area $4\pi r^2$.

Using Eqs.~\eqref{eq:h1 asymptotic},~\eqref{eq:hsinglecovd}, and \eqref{eq:doublecovd}, and the formulas for $\ddG_{ab}$ in the supplemental \pkg{Mathematica} notebook~\cite{2nd-order-notebook}, it is straightforward (but tedious) to show that $\ddG_{nn}\sim r^{-2}$, $\ddG_{n\mb}\sim r^{-3}$ and $\ddG_{\mb\mb}\sim r^{-3}$ if $h_{ll} \sim h_{ln}\sim h_{lm}\sim h_{m\mb} \sim r^{-2}$. By taking this analysis to one further order in $r$, it is possible to show $\ddG_{\mb\mb}\sim r^{-4}$ if $h_{ln}\sim h_{lm}\sim h_{m\mb} \sim r^{-2}$ and $h_{ll} \sim r^{-3}$. Therefore, the source in the Bondi--Sachs gauge behaves as $\S[\ddG_{ab}[h_{ab},h_{ab}]]\sim r^{-4}$, making the retarded integral converge and giving $\psi_{4L}^{(2)}$ the natural behavior $\sim r^{-1}$ of a gravitational wave at \scri. 

We conclude with three notes. First, we observe that since $\psi_{4L}^{(2)}$ and $\psi_{4}^{(2)}$ only differ by terms of order $r^{-2}$, our analysis carries over to $\psi_{4}^{(2)}$ (and to its Campanelli--Lousto source). Second, we observe that the Bondi--Sachs gauge conditions are \emph{not} enforced by standard CCK metric reconstruction. The CCK procedure in the ORG yields a well-behaved solution of the form~\eqref{eq:h1 asymptotic}, but it sets the ``wrong'' tetrad components of $h^{(1)}_{ab}$ to zero. The IRG sets the ``right'' components to zero, but the CCK procedure in the IRG yields a perturbation whose nonzero components blow up as $h^{(1)}_{\mu\nu}\sim r$~\cite{Keidl:2010pm}. Finally, we note an important pitfall of working in a non-Bondi gauge. If $\psi_{4L}^{(2)}$ and $\psi_{4}^{(2)}$ decay like $\sim \ln(\omega r)e^{-i\omega u}/(\omega r)$ rather than like $\sim e^{-i\omega u}/r$, then they cannot be simply substituted into standard formulas for the flux of energy and angular momentum to \scri. In that instance, one would need to first transform to a Bondi-type gauge in order to compute the fluxes.

\section{Conclusion}\label{sec:conclusion}

In this paper, we have advocated for using a \emph{reduced} second-order Teukolsky equation, Eq.~\eqref{eq:pure-2nd-Teuk}. It has two main advantages: its source term is well defined as a distribution in the context of second-order GSF calculations, while the Campanelli--Lousto field equation for the total field $\psi^{(2)}_4$ is less tractable at the particle's position; and after one solves the reduced equation to obtain $\psi^{(2)}_{4L}$, one can readily apply nonvacuum metric reconstruction methods to retrieve the second-order metric perturbation $h^{(2)}_{ab}$. Both of these advantages stem from the fact that the reduced second-order Teukolsky equation is related to the second-order Einstein equation by a linear operation. An additional benefit is that the field variable $\psi^{(2)}_{4L}$ is invariant under infinitesimal tetrad rotations. 

We have also shown that large-$r$, infrared divergences, which plague generic second-order GSF calculations, can be avoided by imposing appropriate gauge conditions on $h^{(1)}_{ab}$. The prime example of such a gauge is the Bondi--Sachs gauge traditionally used to analyse the structure of \scri. The benefits of this gauge apply regardless of whether one solves the reduced equation or the Campanelli--Lousto equation. In the sequel paper~\cite{paperII}, we will present a method of enforcing this gauge condition and constructing a gauge-invariant version of $\psi^{(2)}_{4L}$ that automatically exhibits the correct asymptotically simple behavior. 

Besides our analyses of the source terms near the particle and near \scri, our main new contribution is an explicit expression for the nonlinear source in the reduced second-order Teukolsky equation. This is included in the supplemental \pkg{Mathematica} notebook~\cite{2nd-order-notebook}, which also includes many of the calculations summarized in the paper. In the second sequel paper~\cite{paperIII}, one of us will present a concrete method of calculating spheroidal-harmonic modes of the second-order source. In another associated paper, \cite{coupling}, we present a simplified formula for 
the spin-weighted spherical harmonic modes of the source in the special case of a Schwarzschild background; and in Ref.~\cite{2ndOrderTeukolskyInSchwarzschild} we present an implementation for quasicircular, nonspinning binaries.

A primary goal, in the GSF context, will be the generation of waveforms at first post-adiabatic order (1PA). As discussed in the Introduction, this only requires dissipative, time-antisymmetric effects from the second-order GSF (in addition to all information from first order). In Sec.~\ref{subsec:balance-laws}, we reviewed how, at first order, dissipative effects can be extracted directly from $\psi^{(1)}_4$, without reconstructing the full $h^{(1)}_{ab}$. A similar shortcut might be possible at second order. 

One potential obstacle to finding such a shortcut is that, at first order, for generic, inclined orbits, the shortcut was only possible because the rate of change of the Carter constant could be computed from the radiative piece of the metric perturbation, $h^{{\rm Rad}(1)}_{ab}$. It is unknown whether second-order dissipative effects can likewise be obtained purely from a second-order radiative field. Therefore one might need to implement a complete metric reconstruction in order to compute the 1PA rate of change of the Carter constant. But at least for equatorial orbits, one should be able to entirely bypass $h^{(2)}_{ab}$ by obtaining the 1PA dissipative effects directly from the gravitational-wave fluxes of energy and angular momentum to \scri and down the central black hole's horizon. The fluxes to infinity are readily calculated from $\psi^{(2)}_{4L}$ (or $\psi^{(2)}_{4}$). We are not aware of a practical formula for the fluxes through the horizon beyond leading order, but if a formula for them is derived in terms of $\psi^{(2)}_{4L}$, then one should be able to compute 1PA equatorial evolution directly from $\psi^{(2)}_{4L}$. Recent work for inspiraling scalar charges also suggests it might be possible to compute the rate of change of the Carter constant from true ``fluxes of Carter constant''~\cite{grant2022flux}, which might be extracted from the Weyl scalar without requiring metric reconstruction. 

In considering such shortcuts, one should keep in mind that they do not circumvent the need for metric reconstruction at first order. Information from the complete $h^{(1)}_{ab}$ is needed in almost every aspect of a 1PA calculation: $h^{(1)}_{ab}$ is required to construct the nonlinear source term for  $\psi^{(2)}_{4L}$; the regular field $h^{{\rm R}(1)}_{ab}$ and the GSF it exerts are required in calculations of the material source $T^{(2)}_{ab}$~\cite{Upton,Miller-Pound:2021}; and the complete (conservative plus dissipative) first-order GSF contributes directly to the 1PA phase evolution.

Moreover, in some contexts reconstructing the complete $h^{(2)}_{ab}$ will clearly be required. An example is the calculation of second-order conservative GSF effects, which could be important for improving effective-one-body models~\cite{Bini:2016cje,Bini:2019nra} and improving the accuracy of GSF models for less extreme mass ratios~\cite{Albertini:2022rfe}.



Finally, we indicate again that the formalism here is not limited to GSF calculations. It applies equally well to any second-order perturbative calculation in Kerr spacetime. Issues we have pointed to, such as large-$r$ asymptotics, are also likely to arise in any number of contexts.

\begin{acknowledgments}
We thank Beatrice Bonga and Nicholas Loutrel for independently cross-checking the results in Appendix~\ref{app:1st-order-NP}. We also thank Leor Barack, Eanna Flanagan, and Barry Wardell for helpful discussions. AP acknowledges the support of a Royal Society University Research Fellowship and associated awards, and AP and AS specifically acknowledge the support of a Royal Society University Research Fellowship Enhancement Award. AS additionally acknowledges partial support
from the STFC Consolidated Grant no. ST/V005596/1.

\end{acknowledgments}

\appendix

\section{\label{app:1st-order-NP}First-order Newman--Penrose quantities}

In this appendix we display the first-order perturbations of the tetrad and spin coefficients, which enter into the Campanelli--Lousto formulation of the second-order Teukolsky equation. 

The tetrad legs are expanded as in Eq.~\eqref{eq:tetrad expansion}. As the metric and tetrad satisfy Eq.~\eqref{eq:NPg}, the metric perturbation can be expressed as
\begin{align}\label{eq:NPmetricPert}
  h^{(1)}_{ab} &= -2 l^{(1)}_{(a} n_{b)} -2 n^{(1)}_{(a}l_{b)} + 2 m^{(1)}_{(a} \bar{m}_{b)} \notag\\ 
&\quad + 2 \bar{m}^{(1)}_{(a} m_{b)} + 2 m_{(a} \bar{m}^c h^{(1)}_{b)c} + 2 \bar{m}_{(a} m^c h^{(1)}_{b)c} \notag\\ 
&\quad - 2 l_{(a} n^c h^{(1)}_{b)c} - 2 n_{(a} l^c h^{(1)}_{b)c},
\end{align}
where indices on the tetrad perturbations are lowered with the background metric. A perturbed tetrad which satisfies Eq.~\eqref{eq:NPmetricPert} is given in~\cite{l-c}, reproduced from~\cite{chrztet}. In our sign convention it reads%
\begingroup%
\allowdisplaybreaks%
\begin{subequations}\label{eq:perturbed-tetrad}%
\begin{align}
  l^{(1)a} &= \frac{1}{2} h_{ll} n^a, \\
  n^{(1)a} &= \frac{1}{2} h_{nn} l^{a} + h_{nl} n^a, \\
  m^{(1)a} &= -\frac{1}{2} h_{mm} \bar{m}^{a} - \frac{1}{2} h_{m \bar{m}} m^a  + h_{m l} n^a +  h_{m n} l^a, \\
  \bar{m}^{(1)a} &= -\frac{1}{2} h_{\bar{m}\bar{m}} m^{a} - \frac{1}{2} h_{m \bar{m}} \bar{m}^a  + h_{\bar{m} l} n^a + h_{\bar{m} n} l^a.
\end{align}
\end{subequations}
\endgroup
Note this is not a unique choice of perturbed tetrad; the perturbations transform under the infinitesimal tetrad rotation~\eqref{eq:tetrad rotation}.

Starting from these tetrad perturbations, Campanelli and Lousto also provided the perturbations of the spin coefficients~\cite{l-c}. We were unable to fully reproduce their results. We have performed a full re-derivation and insisted
consistency with the Bianchi identities used in the original derivation. We have
found a slightly corrected full set of first-order spin coefficients, 
\begin{widetext}
{\allowdisplaybreaks
\begin{subequations}
\begin{align}
\kappa^{(1)} &= - \kappa h_{ln} - \tfrac{1}{2} \kappa h_{m\bar{m}} - \tfrac{1}{2} \bar{\kappa} h_{{mm}} - (D -2 \varepsilon - \bar{\rho}) h_{lm} + \sigma h_{l\bar{m}} - (\bar{\alpha} + \beta - \tfrac{1}{2} \bar{\pi} - \tfrac{1}{2}\tau - \tfrac{1}{2} \delta ) h_{ll}, \\
\sigma^{(1)} &= -\tfrac{1}{2} \bar{\lambda} h_{ll} - (\tfrac{1}{2} D - \varepsilon + \bar{\varepsilon} + \tfrac{1}{2} \rho - \tfrac{1}{2} \bar{\rho})h_{mm} - (- \bar{\pi} - \tau)h_{lm},\\
\nu^{(1)} &= \lambda h_{{nm}} - (- \Delta -2 \gamma - \bar{\mu}) h_{n\bar{m}} + \nu h_{ln} - \tfrac{1}{2} \nu h_{m\bar{m}} - \tfrac{1}{2} \bar{\nu} h_{\bar{m}\bar{m}} - (\alpha + \bar{\beta} - \tfrac{1}{2} \pi - \tfrac{1}{2} \bar{\tau} + \tfrac{1}{2} \bar{\delta}) h_{nn}, \\
\lambda^{(1)} &= \lambda h_{ln} - (- \tfrac{1}{2} \Delta - \gamma + \bar{\gamma} + \tfrac{1}{2} \mu - \tfrac{1}{2} \bar{\mu}) h_{\bar{m}\bar{m}} - \tfrac{1}{2} \bar{\sigma} h_{{nn}} - (- \pi - \bar{\tau}) h_{n\bar{m}},\\
\mu^{(1)} &= - (- \tfrac{1}{2} \mu - \tfrac{1}{2} \bar{\mu}) h_{ln} - (- \tfrac{1}{2} \Delta + \tfrac{1}{2} \mu - \tfrac{1}{2} \bar{\mu}) h_{m\bar{m}} - (- \tfrac{1}{2} \delta - \beta - \tfrac{1}{2} \tau)h_{n\bar{m}} \notag\\&\quad- (\tfrac{1}{2}\bar{\delta} + \bar{\beta} - \pi - \tfrac{1}{2} \bar{\tau}) h_{{nm}} + \tfrac{1}{2} \nu h_{{lm}} - \tfrac{1}{2} \bar{\nu} h_{l\bar{m}} - \tfrac{1}{2} \rho h_{{nn}}, \\
\rho^{(1)} &= \tfrac{1}{2} \kappa h_{n\bar{m}} - \tfrac{1}{2} \bar{\kappa} h_{{nm}} - \tfrac{1}{2} \mu h_{{ll}} - (\tfrac{1}{2} \bar{\delta} - \alpha - \tfrac{1}{2} \pi) h_{{lm}} - (\tfrac{1}{2} \rho - \tfrac{1}{2} \bar{\rho}) h_{ln} \notag\\&\quad - (\tfrac{1}{2}D + \tfrac{1}{2} \rho - \tfrac{1}{2} \bar{\rho}) h_{m\bar{m}} - ( - \tfrac{1}{2} \delta + \bar{\alpha} - \tfrac{1}{2} \bar{\pi} - \tau) h_{l\bar{m}},\\
\varepsilon^{(1)} &= \tfrac{1}{4} \kappa h_{n\bar{m}} - \tfrac{1}{4} \bar{\kappa} h_{{nm}} - (- \tfrac{1}{4} \Delta + \tfrac{1}{2} \bar{\gamma} + \tfrac{1}{4} \mu - \tfrac{1}{4} \bar{\mu}) h_{{ll}} - (\tfrac{1}{2}D + \tfrac{1}{4} \rho - \tfrac{1}{4} \bar{\rho}) h_{ln} - (\tfrac{1}{4} \rho - \tfrac{1}{4} \bar{\rho}) h_{m\bar{m}} \notag\\&\quad + \tfrac{1}{4} \sigma h_{\bar{m}\bar{m}} - \tfrac{1}{4} \bar{\sigma} h_{{mm}} - (-\tfrac{1}{4} \delta + \tfrac{1}{2} \bar{\alpha} - \tfrac{1}{4} \bar{\pi} - \tfrac{1}{2} \tau) h_{l\bar{m}} - (\tfrac{1}{4} \bar{\delta} - \tfrac{1}{2} \alpha - \tfrac{3}{4} \pi - \tfrac{1}{2} \bar{\tau}) h_{{lm}}, \\
\pi^{(1)} &= \tfrac{1}{2} \lambda h_{{lm}} - (- \tfrac{1}{2} \Delta + \bar{\gamma} - \tfrac{1}{2} \bar{\mu}) h_{l\bar{m}} - (-\tfrac{1}{2} D - \varepsilon + \tfrac{1}{2} \rho) h_{n\bar{m}} - \tfrac{1}{2} \bar{\sigma} h_{{nm}} \notag\\&\quad + \tfrac{1}{2} \tau h_{\bar{m}\bar{m}} - (\tfrac{1}{2} \bar{\delta} - \tfrac{1}{2} \pi - \tfrac{1}{2} \bar{\tau}) h_{ln} + \tfrac{1}{2} \bar{\tau} h_{m\bar{m}},\\
\tau^{(1)} &= -\tfrac{1}{2} \bar{\lambda} h_{l\bar{m}} - (\tfrac{1}{2} \Delta - \gamma + \tfrac{1}{2} \mu) h_{{lm}} + \tfrac{1}{2} \pi h_{{mm}} + \tfrac{1}{2} \bar{\pi} h_{m\bar{m}} \notag\\&\quad
- (\tfrac{1}{2}D + \bar{\varepsilon} - \tfrac{1}{2} \bar{\rho}) h_{{nm}} + \tfrac{1}{2}\sigma h_{n\bar{m}} - (-\tfrac{1}{2}\delta - \tfrac{1}{2} \bar{\pi} - \tfrac{1}{2} \tau) h_{ln}, \\
\alpha^{(1)} &= -\tfrac{1}{4} \bar{\kappa} h_{{nn}} + \tfrac{3}{4} \lambda h_{{lm}} - (-\tfrac{1}{4} \Delta - \gamma + \tfrac{1}{2} \bar{\gamma} + \tfrac{1}{2} \mu - \tfrac{1}{4} \bar{\mu}) h_{l\bar{m}} - \tfrac{1}{4}\nu h_{{ll}} - (\tfrac{1}{4} D - \tfrac{1}{2} \varepsilon - \tfrac{1}{4} \rho - \tfrac{1}{2} \bar{\rho}) h_{n\bar{m}} \notag\\&\quad- \tfrac{1}{4} \bar{\sigma} h_{{nm}} - (- \tfrac{1}{4}\delta + \tfrac{1}{2} \bar{\alpha} - \tfrac{1}{4} \bar{\pi} - \tfrac{1}{4} \tau) h_{\bar{m}\bar{m}} - (\tfrac{1}{4} \bar{\delta} - \tfrac{1}{4} \pi - \tfrac{1}{4} \bar{\tau}) h_{ln}- (\tfrac{1}{4} \bar{\delta} + \tfrac{1}{2} \alpha - \tfrac{1}{4} \pi - \tfrac{1}{4} \bar{\tau}) h_{m\bar{m}}, \\
\beta^{(1)} &= -\tfrac{1}{4} \kappa h_{{nn}} - \tfrac{1}{4} \bar{\lambda} h_{l\bar{m}} - (-\tfrac{1}{4}\Delta - \tfrac{1}{2} \gamma - \tfrac{1}{4} \mu - \tfrac{1}{2} \bar{\mu}) h_{{lm}} - \tfrac{1}{4} \bar{\nu} h_{{ll}}  - (\tfrac{1}{4} D - \varepsilon + \tfrac{1}{2} \bar{\varepsilon} + \tfrac{1}{2} \rho - \tfrac{1}{4} \bar{\rho}) h_{{nm}} \notag\\&\quad+ \tfrac{3}{4} \sigma h_{n\bar{m}} - (\tfrac{1}{4} \delta - \tfrac{1}{4} \bar{\pi} - \tfrac{1}{4} \tau) h_{ln} - (-\tfrac{1}{4}\delta +\tfrac{1}{2} \beta - \tfrac{1}{4} \bar{\pi} - \tfrac{1}{4} \tau) h_{m\bar{m}} - (\tfrac{1}{4} \bar{\delta} +\tfrac{1}{2} \bar{\beta} - \tfrac{1}{4} \pi - \tfrac{1}{4} \bar{\tau}) h_{{mm}},\\
\gamma^{(1)} &= \tfrac{1}{4} \lambda h_{{mm}} - \tfrac{1}{4} \bar{\lambda} h_{\bar{m}\bar{m}} - (\tfrac{1}{4} \mu - \tfrac{1}{4} \bar{\mu}) h_{m\bar{m}} - (- \gamma + \tfrac{1}{4} \mu - \tfrac{1}{4} \bar{\mu}) h_{ln} + \tfrac{1}{4} \nu h_{{lm}} - \tfrac{1}{4} \bar{\nu} h_{l\bar{m}} \notag\\&\quad- (\tfrac{1}{4} D +\tfrac{1}{2} \bar{\varepsilon} + \tfrac{1}{4} \rho - \tfrac{1}{4} \bar{\rho}) h_{{nn}} - (-\tfrac{1}{4} \delta - \tfrac{1}{2} \beta - \tfrac{1}{2} \bar{\pi} - \tfrac{3}{4} \tau) h_{n\bar{m}} - (\tfrac{1}{4} \bar{\delta}+ \tfrac{1}{2} \bar{\beta} - \tfrac{1}{2} \pi - \tfrac{1}{4} \bar{\tau}) h_{{nm}}.
 \end{align}
\end{subequations}}
\end{widetext}
The calculations made in this appendix are contained in the accompanying \pkg{Mathematica} notebook~\cite{2nd-order-notebook}.

\section{Geroch--Held--Penrose formalism}\label{app:GHP}

The GHP formalism~\cite{geroch1973space}, which is used extensively in the accompanying \pkg{Mathematica} notebook~\cite{2nd-order-notebook}, is a refinement of the NP formalism that is invariant under spin and boost transformations. It is adapted to situations with at least one preferred null direction (e.g., spacetimes with at least one principal null direction), and it effectively halves the formalism's number of equations.

A spin transformation corresponds to 
\beq
m^a\to e^{i\vartheta}m^a,
\eeq
and a boost corresponds to
\beq\label{eq:boost}
l^a\to A l^a\quad \text{and}\quad  n^a\to A^{-1}n^a,
\eeq
where $A>0$ and $0\leq\vartheta<2\pi$ are real numbers. These are the Lorentz transformations that preserve the directions of the null vectors $l^a$ and $n^a$. If one of the vectors is aligned with a principle null direction, then it remains so after the transformation (and therefore, in a Petrov type D spacetime such as Kerr, the two vectors remain aligned with the two principal null directions). 

A tensorial object $\eta$ whose construction involves the tetrad is said to have spin weight $s$ and boost weight $b$ if it transforms as $\eta\to A^{b}e^{is\vartheta}\eta$ under a spin and boost transformation. GHP combined the transformations by defining $\lambda^2 = A e^{i\vartheta}$, such that a quantity $\eta$ of spin weight $s=\frac{1}{2}(p-q)$ and boost weight $b=\frac{1}{2}(p+q)$ transforms as 
\beq
\eta\to \lambda^p\bar\lambda^q\eta. 
\eeq
$\eta$ is then said to have GHP weight $\{p,q\}$. The $\{p,q\}$ weights of the tetrad vectors themselves are $\{1,1\},\{-1,-1\},\{1,-1\}$ and $\{-1,1\}$ for $l^a$, $n^a$, $m^a$, and $\bar{m}^a$, respectively. The product of a quantity of type $\{p,q\}$ and a quantity of type $\{r,s\}$ is of type $\{p+r,q+s\}$.

The alignment of the tetrad (though not the direction of each vector) is also preserved under the interchange of $l^a$ and $n^a$ and under the interchange of $m^a$ and $\mb^a$. GHP denoted the interchange with a prime notation,
\begin{align}
    (l^{a})'=n^a, \ (n^{a})'=l^a, \ (m^{a})'=\mb^a, \ (\mb^{a})'=m^a.
\end{align}
Using this notation, we can express half of the NP spin coefficients as the prime of the other half:
\begin{align}
\kappa^\prime&:=-\nu, \quad \sigma^\prime:=-\lambda, \quad \rho^\prime:=-\mu, \\ \tau^\prime&:=-\pi, \quad \beta^\prime:=-\alpha, \quad \epsilon^\prime:=-\gamma.
\end{align}
Note, under a GHP prime operation, $\{p,q\}\rightarrow \{-p,-q\}$; and under a complex conjugation, $\{p,q\}\rightarrow \{q,p\}$.


It is straightforward to find the GHP weights for eight of the spin coefficients,
\begin{align}
    \kappa:\{3,1\}, \ \sigma:\{3,-1\}, \ \rho:\{1,1\}, \ \tau:\{-1,1\},
\end{align}
and their primes. However, four of the NP spin coefficients do not have well-defined weights: $\epsilon$, $\epsilon^\prime$, $\beta$, and $\beta^\prime$. Additionally, the NP derivatives~\eqref{paradervs} do not have well-defined spin and boost weights. But these quantities can be combined to produce \textit{GHP derivatives} with well-defined weights, given by 
\begin{align}
\thorn = \bm{D}-p\epsilon-q\bar{\epsilon} , \qquad \thorn^\prime  = \bm{\Delta}+p\epsilon^\prime+q\bar{\epsilon}^\prime, \notag \\
\edth  = \bm{\delta}-p\beta+q\bar{\beta} , \qquad \edth^\prime  = \bar{\bm{\delta}}+p\beta^\prime-q\bar{\beta}^\prime, \label{eq:GHPderivatives} 
\end{align}
when acting on a tensor of GHP weight $\{p,q\}$. These derivatives have the following weights: 
\beq
\thorn:\{1,1\}, \ \thorn':\{-1,-1\}, \ \edth:\{1,-1\}, \ \edth':\{-1,1\}.\!\!
\eeq
When written in terms of these derivatives, all equations in the formalism have definite weights.

\section{\label{app:GreenTeuk}GHZ metric reconstruction}

In this appendix, we fill in some of the details of the GHZ reconstruction procedure summarized in Sec.~\ref{sec:reconstruction}. We make heavy use of the GHP notation reviewed in Appendix~\ref{app:GHP}, particularly the GHP prime operation. For clarity, we append labels ``IRG'' and ``ORG'' to the Hertz potentials and corrector tensors. We summarize the scheme at first order and then state the generalization to $n$th order.

Unlike the description in the body of this paper, GHZ's original procedure began from $\psi^{(1)}_{0}={\cal T}'[h^{(1)}_{ab}]$ and worked in an IRG (i.e., $h^{(1)}_{ab}l^b=0$). $\psi^{(1)}_{0}$ satisfies the Teukolsky equation~\eqref{eq:teuk-concise-prime}. Given $\psi^{(1)}_{0}$, the reconstructed metric perturbation is written as
\begin{align}\label{eq:green-metric-recon-IRG}
h^{(1)}_{ab} = 2{\rm Re}(S^{\prime \dagger}_{ab}[\Phi^{(1)}_{\rm IRG}]) + x^{(1)\rm IRG}_{ab}
\end{align}
in terms of a Hertz potential $\Phi^{(1)}_{\rm IRG}$ and a corrector tensor $x^{(1)\rm IRG}_{ab}$. 

The Hertz potential can be obtained from the radial inversion equation\footnote{Our inversion formulas differ by a factor of $-1/2$ relative to the formula in Ref.~\cite{Toomani:2021jlo} due to our mostly positive signature and our convention for the operator ${\cal S}$. Our normalizations agree with Ref.~\cite{Pound-Wardell:2021}. Our formulas for the corrector tensor likewise involve an overall change in sign.}
\beq\label{eq:IRG Hertz}
\frac{1}{4}\thorn^4\bar\Phi^{(1)}_{\rm IRG} = \psi^{(1)}_{0},
\eeq
where the bar denotes complex conjugation. $\thorn$, defined in Eq.~\eqref{eq:GHPderivatives}, is a derivative along principal outgoing null rays, making Eq.~\eqref{eq:IRG Hertz} an ordinary differential equation along those rays. In outgoing null coordinates and the Kinnersley tetrad, $\thorn$ reduces to $\partial_r$.

The corrector tensor in the IRG has nonzero components $x^{(1)\rm IRG}_{m\mb}$, $x^{(1)\rm IRG}_{nm}$, and $x^{(1)\rm IRG}_{nn}$. They can be calculated by solving three hierarchical ordinary differential equations, again along principal outgoing null rays. These equations are
\begin{align}
 \rho^2 \thorn \left[ \frac{\bar\rho}{\rho^3} \thorn \left(\frac{\rho}{\bar\rho} x^{(1)\rm IRG}_{m\mb} \right) \right] = -8\pi T^{(1)}_{ll},
\end{align}
followed by
\begin{multline}
\frac{\rho}{2(\rho+\bar\rho)} 
\thorn\left[ (\rho+\bar\rho)^2\thorn\frac{x^{(1)\rm IRG}_{nm}}{\rho(\rho+\bar\rho)}\right] \\
= -8\pi T^{(1)}_{lm} +{\cal N}[x^{(1)\rm IRG}_{m\mb}],
\end{multline}
and then
\begin{multline}
\frac{1}{2} (\rho+\bar\rho)^2\thorn \left( \frac{1}{\rho+\bar\rho}x^{(1)\rm IRG}_{nn} \right) \\
= -8\pi T^{(1)}_{ln} +{\rm Re}\Bigl({\cal U}[x^{(1)\rm IRG}_{m\mb}]\Bigr) +{\rm Re}\Bigl({\cal V}[x^{(1)\rm IRG}_{nm}]\Bigr),
\end{multline}
where the linear differential operators ${\cal N}$, ${\cal U}$, and ${\cal V}$ are given in Eqs.~(57)--(58) of Ref.~\cite{Toomani:2021jlo}.\footnote{Equation~(58) in Ref.~\cite{Toomani:2021jlo} contains a typo in its second term within curly brackets. $(\tau'-\bar\tau)\edth'$ should be $(\tau'-\bar\tau)\edth$.} The inclusion of a nonzero trace component, $x^{(1)\rm IRG}_{m\mb}$, is essential; the contribution to $h^{(1)}_{ab}$ from the Hertz potential is traceless, and as shown by Ref.~\cite{price2007existence}, it is impossible to put a metric perturbation $h^{(1)}_{ab}$ in a traceless IRG if $T^{(1)}_{ll}\neq0$.

\newpage

In the summary in the body of the paper, instead of the above, we start from $\psi^{(1)}_4$ and reconstruct the metric perturbation in an ORG. This procedure can be derived directly from the one above by applying the GHP prime operation to all equations. $\psi^{(1)}_4$ satisfies the Teukolsky equation~\eqref{eq:teuk-concise}; the reconstructed metric perturbation is given by Eq.~\eqref{eq:green-metric-recon}; the ORG Hertz potential can be obtained from the radial inversion equation
\beq\label{eq:Hertz}
\frac{1}{4}\thorn'^4\bar\Phi^{(1)}_{\rm ORG} = \psi^{(1)}_{4},
\eeq
and the corrector tensor can be found by solving the sequence of ordinary differential equations
\begin{align}
 \rho'^2 \thorn' \left[\frac{\bar\rho'}{\rho'^3} \thorn' \left(\frac{\rho'}{\bar\rho'} x^{(1)\rm ORG}_{m\mb} \right) \right] = -8\pi T^{(1)}_{nn},
\end{align}
followed by 
\begin{multline}
\frac{\rho'}{2(\rho'+\bar\rho')} 
\thorn'\left[ (\rho'+\bar\rho')^2\thorn'\frac{x^{(1)\rm ORG}_{l\mb}}{\rho'(\rho'+\bar\rho')}\right] \\
= -8\pi T^{(1)}_{n\mb} +{\cal N}'[x^{(1)\rm ORG}_{m\mb}],
\end{multline}
and then
\begin{multline}
\frac{1}{2} (\rho'+\bar\rho')^2\thorn' \left( \frac{1}{\rho'+\bar\rho'}x^{(1)\rm ORG}_{ll} \right) \\
= -8\pi T^{(1)}_{nl} +{\rm Re}\Bigl({\cal U}'[x^{(1)\rm ORG}_{m\mb}]\Bigr) +{\rm Re}\Bigl({\cal V}'[x^{(1)\rm ORG}_{l\mb}]\Bigr).
\end{multline}
$\thorn'$, defined in Eq.~\eqref{eq:GHPderivatives}, is a derivative along principal ingoing null rays. In ingoing null coordinates and an appropriate tetrad, $\thorn'$ reduces to $-\partial_r$.

These procedures generalize immediately to $n$th order, starting from $\psi^{(n)}_{0L}:={\cal T}'[h^{(n)}_{ab}]$ or $\psi^{(n)}_{4L}:={\cal T}[h^{(n)}_{ab}]$ and replacing $8\pi T^{(1)}_{ab}$ with the source in the $n$th-order Einstein equation.



\bibliography{bib}

\end{document}